\documentclass[iicol,pdflatex,sn-basic]{sn-jnl}

\usepackage{graphicx}%
\usepackage{multirow}%
\usepackage{amsmath,amssymb,amsfonts}%
\usepackage{amsthm}%
\usepackage{mathrsfs}%
\usepackage[title]{appendix}%
\usepackage[table,xcdraw]{xcolor}%
\usepackage{textcomp}%
\usepackage{manyfoot}%
\usepackage{booktabs}%
\usepackage{algorithm}%
\usepackage{algorithmicx}%
\usepackage{algpseudocode}%
\usepackage{listings}%
\usepackage{enumerate}
\usepackage{enumitem}
\usepackage{mathtools}
\usepackage{tabularx} % For auto-adjusting column widths

\usepackage[utf8]{inputenc}

\usepackage{cleveref}
\crefname{figure}{Fig.}{Figs.}
\crefname{section}{Sec.}{Secs.}
\crefname{equation}{Eq.}{Eqs.}
\crefname{table}{Table}{tables}
\Crefname{figure}{Fig.}{Figs.}
\Crefname{section}{Sec.}{Secs.}
\Crefname{equation}{Eq.}{Eqs.}
\Crefname{table}{Table}{Tables}

\newcommand{\citeasnoun}[1]{\cite{#1}}
\newcommand{\citeasnouns}[1]{\cite{#1}}
\newcommand{\ignore}[1]{}

\newcommand{\R}{\mathbb{R}}

\newcommand{\de}{\,\mathrm{d}}
\newcommand{\bol}{\boldsymbol}

\newcommand{\elf}{\mathbf E}

\newcommand{\lrho}{\boldsymbol{\rho}}  % latent field
\newcommand{\frho}{\boldsymbol{\widetilde\rho}}  % filtered field
\newcommand{\prho}{\boldsymbol{\hat\rho}}  % projected field
\newcommand{\Rrad}{{\widetilde R}}  % filtering radius
\newcommand{\filtk}{{\widetilde k}} % filtering kernel

\newcommand{\nex}{\boldsymbol{x}}
\newcommand{\abs}[1]{\left|#1\right|}
\newcommand{\norm}[1]{\left\lVert#1\right\rVert}
\newcommand{\correction}[1]{{#1}}

\newcommand{\objfun}[1]{
    \scalebox{0.7}{$\begin{pmatrix} \text{obj. =} \\ \text{#1} \end{pmatrix}$}
}
\newcommand{\objfunn}[1]{(\text{#1})
}

\begin{document}

\title{Hyperparameter-free minimum-lengthscale constraints for~topology optimization}

\author*[1]{\fnm{Rodrigo} \sur{Arrieta}}\email{rarrieta@mit.edu}

\author[2]{\fnm{Giuseppe} \sur{Romano}}%\email{romanog@mit.edu}

\author[1]{\fnm{Steven G.} \sur{Johnson}}%\email{iiiauthor@gmail.com}

\affil*[1]{\orgdiv{Department of Mathematics}, \orgname{Massachusetts Institute of Technology}, \orgaddress{\street{77~Massachusetts~Ave.}, \city{Cambridge}, \postcode{02139}, \state{MA}, \country{USA}}}

\affil[2]{\orgdiv{Institute for Soldier Nanotechnologies}, \orgname{Massachusetts Institute of Technology}, \orgaddress{\street{77~Massachusetts~Ave.}, \city{Cambridge}, \postcode{02139}, \state{MA}, \country{USA}}}

\abstract{The geometric constraints of \citeasnoun{zhou_minimum_2015} are a widely used technique in topology/freeform optimization to impose minimum lengthscales for manufacturability. However, its efficiency degrades as design binarization is increased, and it requires heuristic tuning of multiple hyperparameters. In this work, we derive \emph{analytical} hyperparameters from first principles, depending only on the target lengthscale. We present results for both conic and PDE-based filtering schemes, showing that the latter is less robust due to the singularity of its underlying Green’s function. To address this, we also introduce a double-filtering approach to obtain a well-behaved PDE-based filter. Combined with our derived hyperparameters, we obtain a straightforward strategy for enforcing lengthscales using geometric constraints, with minimal hyperparameter tuning. A key enabling factor is the recent subpixel-smoothed projection (SSP) method~\citep{hammond_unifying_2025}, which facilitates the rapidly-converging optimization of almost-everywhere binary designs. The effectiveness of our method is demonstrated for several photonics and heat-transfer inverse-design problems.}


\maketitle

\section{Introduction}\label{sec:introduction}
For practical manufacturing purposes, designs generated via the topology optimization (TO) framework~\citep{bendsoe_generating_1988,jensen_topology_2011,sigmund_topology_2013} (reviewed in \cref{sec:to_overview}) must adhere to fabrication constraints; in particular, designs must typically comply with minimum linewidth and linespacing requirements. A popular technique in density-based TO for imposing these conditions is the geometric-constraints approach, introduced by \citet{zhou_minimum_2015} (reviewed in \cref{sec:geo_constr}). It consists of including two extra constraints into the TO problem: one that imposes a minimum lengthscale on the solid phase and another on the void phase. Both constraints are differentiable and inexpensive to compute, and thus well-suited for gradient-based optimization methods for large-scale nonlinear problems, such as the \correction{conservative convex separable approximation (CCSA)} algorithm~\citep{svanberg_class_2002}. While this approach has proven successful in many works \citep{clausen2015toppoison,ZHAO2018,wangcavity2018,giannini2020topology,hammond_photonic_2021,hammond2022multi,hammond2022phase,hammond2022high,ruan2022inverse,albrechtsen_nanometer-scale_2022,sang2022toward,shang2023inverse,chen_validation_2024,Martinez2024thermooptical,maguirre_omnidirectional_2024,chen2025inverse}, it still has some unsatisfactory aspects. First, there is a delicate interplay between the binarization level of a design---controlled by a projection steepness hyperparameter $\beta$, which is gradually increased during optimization---and the efficacy of the geometric constraints. The lengthscale of a design becomes meaningful only when the design is close to binary, hence the geometric constraints require a high $\beta$ to adequately impose minimum lengthscales. However, this presents a challenge, as standard binarization/projection schemes~\citep{xu_volume_2010,wang_projection_2011} suffer from ill-conditioning and slow convergence as binarization is increased~\citep{hammond_unifying_2025,dunning_automatic_2025}. Consequently, binarization must be limited by a heuristic choice of~$\beta$, hindering the effectiveness of the geometric constraints. Second, the constraints are controlled by several hyperparameters, and guidance on setting these hyperparameters is mostly empirical and subject to cumbersome problem-specific tuning.  For example, to account for non-binary designs, the user must heuristically relax the geometric constraints by means of a threshold $\epsilon$, which is gradually decreased during optimization (see \cref{sec:geo_constr}).  Recently, Hammond et al. introduced a new binarization technique, the subpixel-smoothed projection (SSP)~\citep{hammond_unifying_2025} (\cref{sec:to_overview}), capable of optimizing an \emph{almost-everywhere} binary design (differentiable at $\beta=\infty$) while simultaneously overcoming the issue of slow convergence. Given this new SSP development, can one now revisit the existing geometric constraints algorithm, to obtain a simplified approach with improved performance and minimal reliance on heuristic hyperparameter tuning?

In this paper, we show that the existing geometric constraint algorithm can be greatly simplified and accelerated in conjunction with SSP binarization. In particular, given a desired minimum lengthscale to be imposed on the design, we can now theoretically justify specific choices for \emph{all} geometric constraint hyperparameters (including the filtering radius $\Rrad$, the structural function decay rate $c$, and the constraint threshold $\epsilon$---see \cref{sec:geo_constr} and \cref{tab:summary}), leading to a scheme that requires minimal hyperparameter tuning and straightforwardly achieves rapid convergence. Furthermore, our choice of hyperparameters leads to resolution-invariant constraints, which do not require further adjustment if resolution is increased. Initially, we focus on the case where a conic filtering scheme is used in the TO scheme (\cref{sec:param_deriv}), but we also provide the analysis and recommended hyperparameters when a PDE filter~\citep{lazarov_filters_2011} is employed instead (\cref{sec:pde_bipde_filters} and \cref{sec:closed_form_densities_pde}). While the PDE filter is convenient in unstructured meshes, we show that its singular Green's function makes the geometric constraints much less robust in the presence of small features. To address this difficulty, we propose a simple solution: applying the PDE filter twice. This ``bi-PDE'' filter scheme is analyzed in \cref{sec:pde_bipde_filters} and \cref{sec:closed_form_densities_bipde}.

Our resulting TO algorithm (\cref{sec:strategy}) consists of two stages: First, a design is optimized \emph{without imposing} geometric constraints while rapidly increasing SSP's $\beta$ up to $\beta=\infty$. Then, when the objective function reaches a satisfactory value, the geometric constraints are imposed (at $\beta=\infty$) with hyperparameter values \emph{determined analytically} from the desired lengthscale, and the design is evolved until the constraints become feasible and the objective function attains an acceptable value (typically, this requires fewer iterations than the initial unconstrained optimization). 
Because geometric constraints are only a proxy for
the true fabrication constraints~\citep{zhou_topology_2014,zheng_manufacturing2023}, and in any case
the precise definition of ``minimum lengthscale'' for
freeform geometries is inherently ambiguous (especially
at the scale of the spatial discretization)~\citep{chen_validation_2024}, some \textit{a~posteriori} threshold tuning is occasionally still required to tweak the resulting geometry. To demonstrate the applicability of our approach, we present four photonics TO examples and one heat-transport TO example in \cref{sec:results}, all of them showing good agreement between the obtained and imposed lengthscales. Our strategy and selection of hyperparameters for the three filter cases (conic, PDE, and bi-PDE filters) is summarized in \cref{tab:summary}. Since the resulting algorithm is simply a streamlined version of an existing geometric-constraints method, we believe that it will be attractive to adopt for many applications of TO (in conjunction with SSP binarization), as well as raising new questions for future research (\cref{sec:conclusions}).

Several alternative approaches to impose minimum lengthscales on TO designs have been proposed. The two-phase Heaviside projection method (HPM) \citep{guest_achieving_2004,guest_topology_2009,carstensen_projection_2018} is able to impose minimum lengthscales on both solid and void phases---without additional constraints and with minimal hyperparameter tuning; however, the resulting designs are not fully binarized. Gray pixels values are penalized using, e.g., the solid isotropic material with penalization (SIMP) \citep{bendsoe_optimal_1989} or the rational material penalization (RAMP) \citep{stolpe_alternative_2001} methods, which may still leave some gray pixels that could affect the imposed lengthscales. \cite{sigmund_morphology-based_2007} employs differentiable morphology-based filters---such as erosion, dilation, opening, and closing operators---in conjunction with the SIMP method to impose minimum lengthscales on one or two phases, but again, a resulting binary design is not guaranteed. A robust TO formulation was proposed by \cite{wang_projection_2011}, which accounts for over/under-etch variability while enforcing a quasi-minimum lengthscale. This is achieved by performing a worst-case optimization across three designs---nominal, eroded, and dilated---obtained either by tuning a projection parameter \citep{qian_topological_2013} or, alternatively, through morphological operations \citep{hammond_photonic_2021}. For the method to be valid, the three designs must share the same topology (in \cite{hammond_photonic_2021}, this is enforced via geometric constraints). In \cite{zhang_explicit_2014}, the structural skeleton of the physical design is extracted and minimum lengthscales are imposed on the solid phase by enforcing pixels within a neighborhood of the skeleton to be solid; however, they do not present exact constraint derivatives: their gradient computation neglects changes in the skeleton. \cite{hagg_minimum_2018} show that designs possessing a minimum lengthscale in the solid (void) phase are morphologically open (closed). Based on this fact, the authors propose a modified SIMP method for minimum compliance problems using differentiable open and close operators \citep{svanberg_density_2013} to impose minimum lengthscale on both phases. In \cite{schubert_inverse_2022}, the authors introduce an always-feasible TO method that guarantees a binary and lengthscale-compliant design. The method uses a non-differentiable conditional generator of feasible designs in conjunction with a straight-through gradient estimator. The problem of designing a lengthscale-constrained TO design is then cast as an unconstrained stochastic gradient-optimization problem. In \cite{li_explicit_2023}, a condition to obtain lengthscale-compliant designs is derived---which is somewhat analogous to the morphological condition given in \cite{hagg_minimum_2018}. From this, two differentiable constraints are proposed to enforce minimum lengthscales on both phases. These constraints introduce new hyperparameters into the TO problem; however, the authors provide a heuristic choice and a continuation scheme. In general, methods based on morphological filters require a regularization parameter to make these filters differentiable, and this parameter introduces a tradeoff between binarization and convergence rate analogous to $\beta$, and may also require additional penalty terms to discourage grayscale regions. On the other hand, methods based on non-differentiable constraints or approximate gradients may limit the applicability of optimization algorithms.
\begin{table*}[ht]
		\centering
            \caption{Summary of recommendations and hyperparameters for the conic, PDE, and Bi-PDE filters.}
		\renewcommand{\arraystretch}{1.3}  % increase row height
            % First table
            \scalebox{0.7}{
		\begin{tabular}{@{}>{\centering}p{1.8cm}p{5.5cm}p{3cm}p{3cm}p{3cm}@{}}
			\toprule
			\multicolumn{2}{c@{}}{Geometric constraints variables}&Conic filter&PDE filter&Bi-PDE filter\\
			\multicolumn{2}{c@{}}{(\cref{sec:geo_constr})} & (\cref{sec:param_deriv}) & (\cref{sec:closed_form_densities_pde}) & (\cref{sec:closed_form_densities_bipde})\\
			\midrule
			$\ell_t$ & target lengthscale 
			& user-defined & (same) & (same)  \\
			$\Rrad$ & filter radius 
			& $\ell_t \:\left(\text{or } \in \left[\tfrac{2}{3}\ell_t, 4\ell_t\right]\right)$ & (same) & (same) \\
			$c$ & decay rate 
			& $64\Rrad^2$ & $10\Rrad^2$ & $64\Rrad^2$ \\
			$\epsilon$ & constraint threshold
			& $10^{-8}$ & $\left(\gamma\left(\tfrac{\ell_t}{\Rrad}\right)\right)^{-3}\cdot 10^{-6}$ & $\left(\gamma\left(\tfrac{\ell_t}{\Rrad}\right)\right)^{-3}\cdot 10^{-8}$ \\
			$\gamma\left(\tfrac{\ell_t}{\Rrad}\right)$ & correction-factor function
			& -- & \cref{eq:gamma_pde} & \cref{eq:gamma_bipde} \\
			$\eta_{e,d}\left(\tfrac{\ell_t}{\Rrad}\right)$ & solid/void-threshold functions
			& \cref{eq:eta_e} & \cref{eq:eta_e_pdefilter} & \cref{eq:eta_e_bipdefilter} \\
			\botrule
		\end{tabular}}
            %}\\
            %\vspace{0.5cm}
            % Second table
            \scalebox{0.7}{
		\begin{tabular}{l c p{7cm}}
			\toprule
			{Optimization aspect} &{Specification} &{Notes} \\
			\midrule
			Objective function $f$ & $f \in [1,100]$ & suggested scaling (\cref{sec:strategy}) \\
			Constraint function(s) & $g_\text{geo-c} := \tfrac{g(\rho,\ell_t)}{\epsilon}-1 \leq 0$ & scaled by $1/\epsilon$ (\cref{sec:strategy}) \\
			Projection function & subpixel smoothed projection (SSP)\ignore{~\cite{hammond_unifying_2025}} & allows $\beta=\infty$ \\
			Optimizer & CCSAQ\ignore{~\cite{svanberg_class_2002}} & many alternatives, e.g.~Ipopt\ignore{~\cite{wachter2006implementation}} \\
			Strategy & \multicolumn{2}{p{11cm}}{See \cref{sec:strategy} for details.\begin{enumerate}[leftmargin=*, label=(\roman*), itemsep=0pt]
					\item Optimize an \emph{unconstrained} design while ramping $\beta$ up to $\infty$, until objective function is sufficiently low.
					\item Impose geometric constraints with recommended hyperparameters and $\beta=\infty$; optimize \emph{constrained} design until some stopping criteria is met (e.g., constraints are feasible and obj. fun. is sufficiently low).
					\item In some instances, the imposed minimum lengthscales may be close to, but still less than, the desired target lengthscales. To further increase the imposed lengthscales, reduce $\epsilon$ and continue optimizing.
			\end{enumerate}} \\
			\botrule
		\end{tabular}
            }
		\label{tab:summary}
	\end{table*}

\section{TO overview}\label{sec:to_overview}
In this section we briefly review a general TO design problem. The goal is to find a material distribution in a design region $\Omega \subset \R^d$, with $d=2$ or $d=3$, parametrized by a \emph{latent density field} $\lrho(\nex)$ at each point $\nex\in\Omega$, such that an objective function $f$ is minimized. A typical formulation reads:
\begin{equation} \label{eq:to_problem}
\begin{aligned}
\min_{\lrho} \quad & f(\bol u)  \\
\text{s. t.} \quad & \mathcal{K}(\bol u, \lrho)=0 \\
& g_k(\lrho) \leq 0 & k\in\{1,2,\dots,K\}\qquad \\
& 0\leq\lrho\leq 1 
\end{aligned}
\end{equation}
where $f$ is the objective function, $\bol u$ is the system response and solution of $K(\bol u, \lrho)$, where $\mathcal{K}$ models the physics of the problem and is parametrized by the latent density $\lrho$, and $g_k$ is the $k$th constraint. 

For a photonics TO problem, a common formulation reads:
\begin{equation} \label{eq:photon_to_problem}
\begin{aligned}
\min_{\lrho} \quad & f(\bol \elf_1, \elf_2, \dots, \elf_M) \\
\text{s. t.} \quad &  \nabla\times\tfrac{1}{\mu_0}\nabla\times\elf_m-\omega_m^2\epsilon_0\bol\epsilon_r(\lrho)\elf_m=-i\omega_m \bol J_m \quad  \\
& g_k(\lrho) \leq 0  \\
& 0\leq\lrho\leq 1
\end{aligned}
\end{equation}
where each electric-field response $\elf_m$,~$m\in\{1,2,\dots,M\}$,  depends on a given angular frequency $\omega_m$ and/or current density $\bol J_m$, $\mu_0$ is the vacuum permeability, $\epsilon_0$ is the vacuum permittivity, and $\bol\epsilon_r(\lrho)$ is the point-wise relative permittivity parametrized by the latent density~$\lrho$.

A well-established strategy to parametrize the material distribution is the so-called three-density scheme~\citep{sigmund_topology_2013,wang_projection_2011}. First, to regularize the TO problem and prevent the occurrence of checkerboard patterns and mesh-dependent solutions~\citep{bourdin_filters_2001,bruns_topology_2001}, a low-pass filter is applied to the latent density $\lrho$, resulting in the \emph{filtered density field} $\frho$. There are two widely employed alternatives for filtering. A first alternative is to use a conic filter of radius $\Rrad$, given by
\begin{align}\label{eq:conic_filter}
    \filtk_\Rrad(\nex) = 
    \begin{cases}
        a\left(1-\tfrac{\abs{\nex}}{\Rrad}\right), \quad & \abs{\nex}\leq \Rrad,\\
        0, \quad & \text{otherwise,}
    \end{cases}
\end{align}
where the constant $a$ is chosen such that $\int \filtk_\Rrad = 1$. The filtered field is then obtained as the convolution $\frho = \lrho \ast \filtk_\Rrad$, where $\ast$ denotes 2D or 3D convolution, depending on the problem. A second alternative---convenient for unstructured meshes and finite-element discretizations---is to use a PDE filter approach~\citep{lazarov_filters_2011}, where the filtered density $\frho$ is obtained as the solution of the following ``modified Helmholtz'' equation (essentially, an exponentially damped diffusion process): 
\begin{align}\label{eq:pdefilter_pde}
    \left[-\left(\frac{\Rrad}{2\sqrt{3}}\right)^2\Delta+1\right]\frho = \lrho,
\end{align}
with Neumann (or Dirichlet~\citep{lazarov_robust_2011,qian_topological_2013}) boundary conditions, where $\Rrad$ plays a similar role as the conic filter radius. \correction{Despite its convenience,} we have found that the PDE filter yields a filtered density with large spatial derivatives \correction{$\norm{\nabla_{\nex}\frho(\nex)}$}---\correction{which are attributed to} the singularity and unbounded derivatives of the 2D Green's function of the PDE---and this causes difficulties with the application of geometric constraints (see details in \cref{sec:pde_bipde_filters}). In this work we propose a third alternative---termed ``bi-PDE'' filtering---which simply involves applying the PDE filter twice in succession, resulting in a well-behaved filtered density that is better suited for the geometric constraints. Throughout the rest of the paper we focus our attention on the conic filter case; the particular analyses for the PDE and bi-PDE filter cases are given in \cref{sec:pde_bipde_filters}.

Once the filtered density $\frho$ is obtained, a projection function is applied to obtain a binary structure. This result in a \emph{projected density field} $\prho$, which determines the \emph{physical} distribution of material or void in the design region. Typically, the following smooth tanh-projection~\citep{xu_volume_2010,wang_projection_2011} is employed:
\begin{align}\label{eq:tanh_proj}
    \prho = P_{\beta,\eta}(\frho) =\frac{\tanh(\beta\eta)+\tanh(\beta(\frho-\eta))}{\tanh(\beta\eta)+\tanh(\beta(1-\eta))},
\end{align}
where $\eta$ is the threshold (usually taken as $\eta=1/2$; values of $\frho$ greater than $\eta$ are projected towards one, lower values towards zero) and $\beta > 0$ is the steepness hyperparameter ($\beta=0$ corresponds to the identity mapping $\prho=\frho$, whereas $\beta=\infty$ corresponds to a Heaviside step function $\prho=H(\frho-\eta)$). During optimization $\beta$ is gradually increased from lower values---allowing smooth transitions of topologies---towards higher values to enforce binarization. Finally, the material relative permittivities, for a photonics TO problem, are linearly interpolated \citep{andkjaer_inverse_2014} employing the projected density $\prho$ as:
\begin{align}
    \bol\epsilon_r(\prho) = \epsilon_\text{min}+\prho(\epsilon_\text{max}-\epsilon_\text{min}),
\end{align}
where $\epsilon_\text{min}$ and $\epsilon_\text{max}$ correspond to the relative permittivities of the void and solid region, respectively. Depending on the problem, other interpolating strategies might be employed, e.g., linearly interpolating the inverse of the permittivities~\citep{wadbro_topology_2015} or linearly interpolating the refraction indices~\citep{christiansen_non-linear_2019}.

\correction{
\subsection{Subpixel-smoothed projection}\label{sec:ssp}
It has been noted that the tanh-projection (\cref{eq:tanh_proj}) suffers from ill-conditioning as~$\beta\rightarrow\infty$ \citep{guest_eliminating_2011,dunning_automatic_2025,hammond_unifying_2025}. In the $\beta \to \infty$ regime, the sensitivities of $\hat{\rho}$ approach zero everywhere except at the interfaces where $\frho = \eta$. At these interfaces (a set of measure zero), the sensitivities diverge. Furthermore, even for finite but large~$\beta$, the second derivatives of $\prho$ become large, resulting in an ill-conditioned Hessian that slows down optimization convergence~\citep{nocedal2006numerical,Bertsekas2016}. In practice, to overcome the slow convergence, the $\beta$ step size must be manually tuned during optimization to obtain good designs in reasonable time, and $\beta$ must be capped to a predefined maximum value (often around $\beta\sim 60$), which leads to designs that are not fully binary. (Additional penalties may be used to enforce further binarization, but they come with their own hyperparameters~\citep{Jensen05,Hansen_2024,chen2025inverse}.)
}

In this work, instead of using the tanh-projection as projection function, we employ the recently introduced subpixel-smoothed projection (SSP) technique~\citep{hammond_unifying_2025}. \correction{SSP is an attractive drop-in alternative that: 1) overcomes the finite-$\beta$ slow convergence, 2) leads to a projection that is twice differentiable with \emph{bounded} second derivatives, and 3) allows the optimization of \emph{almost}-everywhere binary designs (binary except for a 1-pixel gray layer adjacent to interfaces)}. In brief, SSP consists of replacing \cref{eq:tanh_proj} with a function~$\prho = \operatorname{SSP}_{\beta,\eta}\left(\frho,\lVert\nabla\frho\rVert\right)$, which remains twice differentiable even at $\beta=\infty$---except at changes in topology, where it falls back to the tanh-projection---while being \emph{almost}-everywhere binary. The key idea is that the inclusion of $\nabla \frho$ information provides knowledge of a local neighborhood of $\frho$, not just a single point, allowing for a smooth dependence on the location of a nearby interface. To smoothly optimize changes in topology, one still optimizes at $\beta$ that increases in steps, but now $\beta$ can be increased much more rapidly, set to $\infty$ as soon as the topology stabilizes. In the present work, the lengthscale constraints are only imposed in this final $\beta = \infty$ epoch, \correction{using analytically derived hyperparameters (\cref{sec:param_deriv_general})}, allowing us to rely on almost-everywhere binarization without sacrificing optimization convergence rate. \correction{We refer the reader to~\cite{hammond_unifying_2025} for more technical details on SSP.}

\section{Geometric constraints}\label{sec:geo_constr}

\begin{figure*}[h!]
\centering
\includegraphics[width=0.8\textwidth]{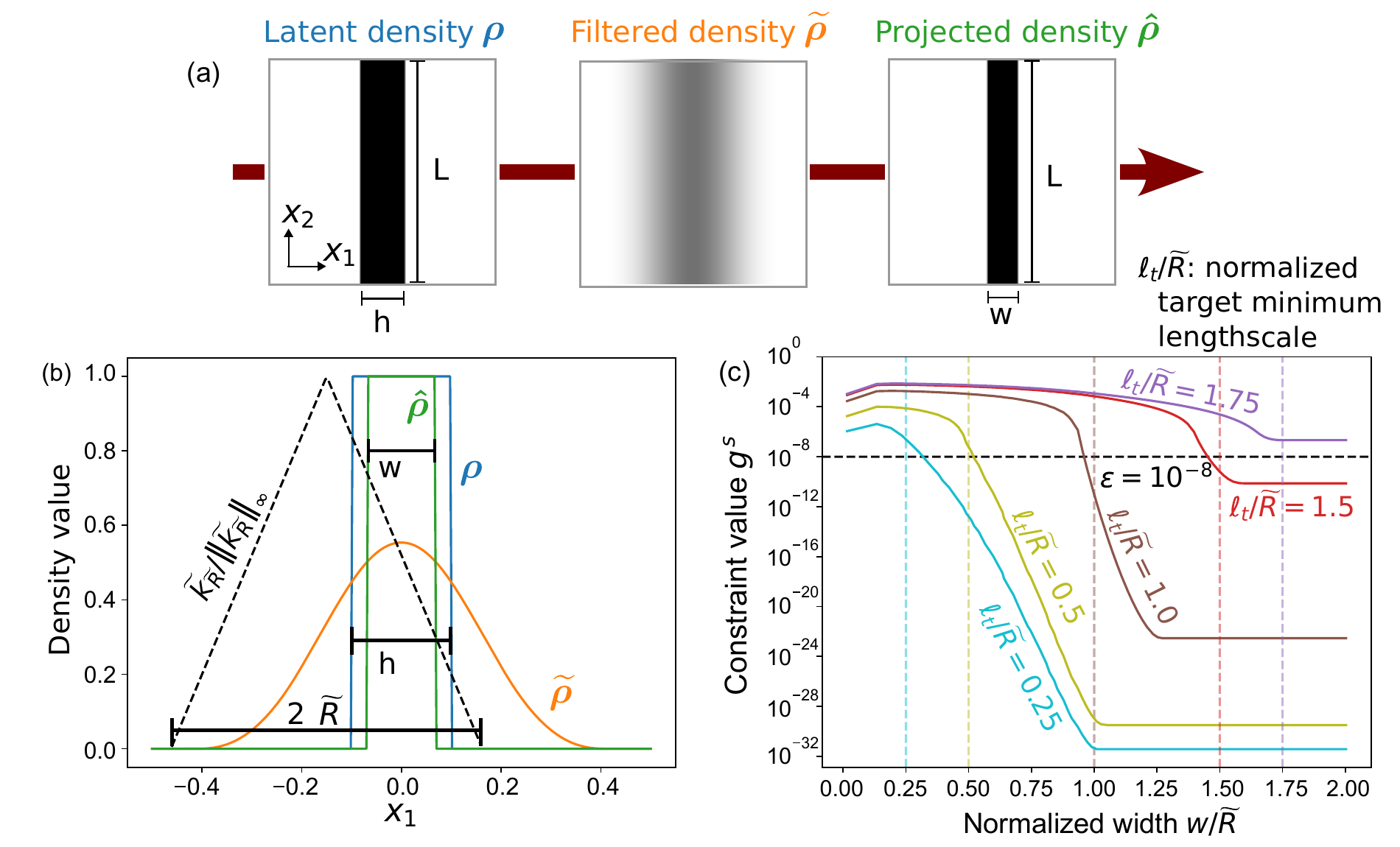} 
\caption{Simple 2D example. (a) Latent density $\lrho$ consisting of a binary strip of width $h$ and length $L \gg h$, which is filtered with a conic filter of radius $\Rrad$ (i.e., $\frho=\lrho\ast{\filtk}_\Rrad$), and then projected with $\beta=\infty$ (i.e., $\prho = H(\frho-0.5)$), resulting in a binary strip of width $w\leq h$ and length $L$. (b) Horizontal cross-section of $\lrho$, $\frho$, $\prho$. A scaled and shifted conic filter kernel $\tilde k$ is depicted as a black dashed line. (c) Value of the solid geometric constraint $g^s$ (\cref{eq:gs_constr}) as a function of the (normalized) physical width $w/\Rrad$, for various (normalized) target lengthscales $\ell_t/\Rrad$ represented with different colors. The vertical dashed lines indicate the value of $\ell_t/\Rrad$; the solid lines indicate the constraint value. We used hyperparameters $c=64\Rrad^2$ and $\epsilon=10^{-8}$ (\cref{tab:summary}). The square design region of size $L\times L$ was discretized with $1024\times1024$ pixels, with $L=1$ and $\Rrad=0.3125$.} 
\label{fig:line_density}
\end{figure*}

Here, we review the geometric constraints introduced in~\citeasnoun{zhou_minimum_2015}, which impose a minimum lengthscales in both the solid and void regions. To impose a minimum target lengthscale $\ell_t$ on the solid region, the following constraint is employed:
\begin{align}\label{eq:gs_constr}
    g^s(\lrho,\ell_t) = \frac{1}{N}\sum_{i=1}^N \bol I_i^s \left[\min\{\frho_i-\eta_e\left(\tfrac{\ell_t}{\Rrad}\right), 0\}\right]^2 \leq \epsilon,
\end{align}
where $N$ is the total number of pixels, $\bol I^s$ is the \emph{solid structural function}, $\eta_e$ is a threshold that depends on the fraction $\ell_t/\Rrad$, and the constraint threshold $\epsilon$ is a small number. The solid structural function $\bol I^s$ indicates the $\emph{inflection region}$ of the solid, given by $\Omega_s = \{\nex \in \Omega \,|\, \prho(\nex)=1\text{ and } \nabla\frho(\nex) = 0\}$, and decays to zero outside of this region. This is achieved with
\begin{align}\label{eq:is_func}
    \bol I^s = \prho \cdot \exp(-c\abs{\nabla\frho}^2),
\end{align}
where $c$ is a hyperparameter that determines the decay rate of $\bol I^s$ outside of the inflection region. On the other hand, if a conic filter of radius $\Rrad$ is used, the \emph{solid threshold function} $\eta_e$ is given by~\citep{qian_topological_2013}:
\begin{align}\label{eq:eta_e}
    \eta_e\left(\tfrac{\ell_t}{\Rrad}\right) = 
    \begin{cases}
    \frac{1}{4}\left(\frac{\ell_t}{\Rrad}\right)^2 + \frac{1}{2}, \quad &\frac{l_t}{\Rrad}\in[0,1],\\
    -\frac{1}{4}\left(\frac{\ell_t}{\Rrad}\right)^2+\frac{\ell_t}{\Rrad}, \quad &\frac{\ell_t}{\Rrad}\in[1,2],\\
    1,\quad  &\frac{\ell_t}{\Rrad}\in[2,\infty).
    \end{cases}
\end{align}
If a PDE or bi-PDE filter is used instead, the analytical expression of $\eta_e$ changes: see \cref{eq:eta_e_pdefilter,eq:eta_e_bipdefilter}, respectively. Analogously, to impose a minimum target lengthscale $\ell_t$ on the void region, the following constrained is imposed:
\begin{align}\label{eq:gv_constr}
    g^v(\lrho,\ell_t) = \frac{1}{N}\sum_{i=1}^N \bol I_i^v \left[\min\{\eta_d\left(\tfrac{\ell_t}{\Rrad}\right)-\frho_i, 0\}\right]^2 \leq \epsilon,
\end{align}
where $\bol I^v$ is the \emph{void structural function}, which identifies the inflection of the void, i.e., $\Omega_v = \{\nex \in \Omega \,|\, \prho(\nex)=0\text{ and } \nabla\frho(\nex) = 0\}$, and is given by
\begin{align}\label{eq:iv_func}
    \bol I^v = (1-\prho) \cdot \exp(-c\abs{\nabla\frho}^2),
\end{align}
whereas, using a conic filter, the \emph{void threshold function} $\eta_d$ reads~\citep{qian_topological_2013}:
\begin{align}\label{eq:eta_d}
    \eta_d\left(\tfrac{\ell_t}{\Rrad}\right) = 
    \begin{cases}
    \frac{1}{2}-\frac{1}{4}\left(\frac{\ell_t}{\Rrad}\right)^2, \quad &\frac{\ell_t}{\Rrad}\in[0,1],\\
    1+\frac{1}{4}\left(\frac{\ell_t}{\Rrad}\right)^2-\frac{l_t}{\Rrad}, \quad &\frac{\ell_t}{\Rrad}\in[1,2],\\
    0,\quad  &\frac{\ell_t}{\Rrad}\in[2,\infty).
    \end{cases}
\end{align}
It is illustrative to qualitatively examine how the constraint value varies as the minimum lengthscale of the projected density $\prho$ changes. Consider the 2D example presented in \cref{fig:line_density}a, in which a latent density $\lrho$, consisting of a binary strip of width $h$ and length $L\gg h$, is filtered with a conic filter of radius $\Rrad$ and then projected (using $\beta=\infty$), resulting in another binary strip of width $w=w(h,\Rrad)$ and same length $L$, where $w$ is the minimum lengthscale of the design. In \cref{fig:line_density}b we show the cross-sections of the densities $\lrho$, $\frho$, and $\lrho$, alongside a (scaled and shifted) cross-section of the conic filter $\filtk_\Rrad$. In \cref{fig:line_density}c we show the value of the solid constraint $g^s(\lrho,\ell_t)$, computed with \cref{eq:gs_constr}, as a function of the \emph{physical} width $w$, for various target minimum lengthscales $\ell_t$. Each target $\ell_t$ is denoted with a different color: vertical dashed lines indicate the (normalized) value of $\ell_t/\Rrad$, whereas solid lines indicate the constraint value. In this example, we use the suggested hyperparameter values of $c=64\Rrad$ and $\epsilon=10^{-8}$ for the conic filter, shown in \cref{tab:summary} and derived in \cref{sec:param_deriv}, to show the correct behavior of the geometric constraint. Examining the $\ell_t/ \Rrad=1$ case, we observe that when the physical lengthscale falls below the target lengthscale (i.e., $w<\ell_t$, indicating a lengthscale violation) the constraint~$g^s$ attains an almost-constant value ($\sim 10^{-4}$ in this case). As $w$ approaches $\ell_t$, $g^s$ decays rapidly and smoothly (note the log scale), and finally stabilizes at a small constant value ($\sim 10^{-22}$ in this case) when $w>\ell_t$ (i.e., the minimum lengthscale is satisfied). Ideally, when $w=\ell_t$, the value of the constraint should equal the threshold (i.e., $g^s=\epsilon$) since~$\epsilon$ determines the boundary between the allowed and excluded lengthscales. This is not the case here, however, due to numerical errors and other simplifications in our derivation (\cref{sec:param_deriv}). Nevertheless, the true intersection has a small error $\Delta w$ less than $\tfrac{1}{16}\Rrad$. Thus, this choice of parameter is adequate for imposing a minimum lengthscale of $\ell_t=\Rrad$, with some acceptable error. Importantly, a higher value of~$c$ results in a sharper transition at $w=\ell_t$, but it could also lead to numerical inaccuracies. Our relatively high $c$ value ensures a sharp transition while avoiding numerical inaccuracies, resulting in a robust constraint with respect to variations of~$\epsilon$ in the same order of magnitude. Lastly, note that, for the same hyperparameters values, the constraint exhibits a similar behavior for other ratios of $\ell_t/R$, approximately in the range $[0.25, 1.5]$---or equivalently, $\Rrad \in \left[\tfrac{2}{3}\ell_t,4\ell_t \right]$. On the other hand, higher ratios, such as $\ell_t/R=1.75$, do not intersect with $\epsilon$, whereas lower ratios become overly sensitive to variations of~$\epsilon$, and thus they are not adequate for imposing minimum lengthscales. 

\section{Hyperparameter derivations}\label{sec:param_deriv_general}

In this section we derive appropriate values for the hyperparameters $\Rrad$, $c$ and $\epsilon$, in the limit of infinite resolution and full binarization ($\beta=\infty$), using a first-principles analysis of stripe/slab geometries.  (The quasi-empirical hyperparameter suggestions from previous works, which typically required problem-specific tuning, are summarized in more detail in the Supplementary Information.)

\subsection{Conic filters}\label{sec:param_deriv}

For simplicity, we focus our attention on the solid constraint $g^s(\lrho,\ell_t)$ of \cref{eq:gs_constr}, which depends on the target lengthscale $\ell_t$ to be imposed. As a first requirement, the constraint must be resolution-invariant, i.e., for a fixed design $\lrho$, the value of $g^s-\epsilon$ must be (approximately) the same irrespective of the sampling resolution (at least in the limit of high enough resolution). and, in fact, the constraint $g^s$ must converge to a unique value at infinite resolution. A sufficient condition for this is that $c$, $\Rrad$, and $\epsilon$ do not depend on the resolution $\Delta x$. As a result, the sum in the constraint can be seen as a Riemann sum, and thus the infinite-resolution constraint value is given by replacing the sum in $g^s$ with an integral, i.e., $\frac{1}{N}\sum_i\rightarrow\frac{1}{\abs{\Omega}}\int_\Omega$, where $\abs{\Omega}$ is the area/volume of the design region $\Omega$. 

We now apply this idea to obtain a closed-form expression for $\epsilon$ in the limit of infinite resolution. Note that, if the projected density $\prho$ contains features whose minimum lengthscale is less than $\ell_t$, then the constraint must be violated, i.e., $g^s(\lrho,\ell_t) > \epsilon$. Conversely, if the minimum lengthscale is greater than $\ell_t$, the constraint must be feasible, implying $g^s(\lrho,\ell_t) < \epsilon$. Therefore, by continuity of $g^s$, if the minimum lengthscale of $\prho$ is exactly $\ell_t$, then $g^s(\lrho,\ell_t) = \epsilon$. It follows that the value of $\epsilon$, in the limit of infinite resolution, is given by evaluating $g^s(\lrho,\ell_t)$ when the projected density $\prho$ has a minimum lengthscale of $\ell_t$. This results in a formula for $\epsilon$ involving the decay rate $c$, the filter radius $\Rrad$, and the imposed lengthscale $\ell_t$ (we eliminate the $\ell_t$ dependence by appropriately choosing $\Rrad$ and $c$, see \cref{eq:epsilon_apprx} below).

Following the same approach as \citeasnouns{qian_topological_2013,trillet_analytical_2021}, we consider a simplified 1D setting where the latent density $\lrho$ consists of a binary bump feature of width $h$, given by
\begin{align}\label{eq:rho_strip}
    \lrho(x) &= \operatorname{rect}\left(\frac{x}{h}\right) = 
    \begin{cases}
        1, \quad&  \abs{\frac{x}{h}} \leq \frac{1}{2}\\
        0, \quad&\text{otherwise}
    \end{cases}
\end{align}
for $x\in\R$, where $\operatorname{rect}$ is the rectangular function. The setting can by extended to 2D by considering a 2D binary strip instead, of width $h$ and length $L\gg h$, as shown in \cref{fig:line_density}a, where a horizontal cross-section of the strip corresponds to the 1D latent density. The same applies to 3D, where we consider instead a binary slab of thickness $h$, length $L$ and width also $L$ (for simplicity). The simplified setting is not general: in practice, the latent density $\lrho$ does not have to be binary and its features can have a combination of different geometries, not necessarily stripes or slabs, and this alters the final expression for $\epsilon$. Unfortunately, the geometric constraints apply, by construction, only to the simplified scenario, but in practice, they are also capable of imposing lengthscales on more general designs. Because of this, the final calculation of $\epsilon$ is not definitive; in some cases it might require some problem-dependent adjustments. 

Continuing with the calculation of $\epsilon$, we filter the 1D latent density $\lrho$ using a 1D conic filter (\cref{eq:conic_filter}) of radius $\Rrad$ and normalization $a=1/R$, yielding a filtered density $\frho$, which is then projected with $\beta=\infty$, resulting in a projected density $\prho$ consisting of a binary bump of width $w$, as shown in \cref{fig:line_density}b. The closed-form expressions of $\frho$, $\prho$, and $w$ are given in \cref{sec:closed_form_densities}. Since $w$ is the minimum lengthscale of $\prho$, we set the imposed lengthscale as $\ell_t=w$. We still have the freedom of choosing the filter radius $\Rrad$ to any value, as long as $\ell_t < 2\Rrad$ (higher values are indistinguishable, see \cref{eq:eta_e}), but it turns out that considering $\Rrad = h$ is sufficient; any other choice of $\Rrad$ with $\Rrad \approx h$ leads to essentially the same expression of $\epsilon$ (see \cref{sec:closed_form_densities}, where we derive analogous expression for $\epsilon$ for other choices of $\Rrad$; we also justify this in \cref{sec:closed_form_densities_pde}).

The choice $\Rrad=h$ leads to $\Rrad=h=w=\ell_t$ according to \cref{eq:h_to_l,eq:l_to_h}, and to $\eta_e=0.75$, according to \cref{eq:eta_e}. Now, using the closed-form expressions of the filtered density $\frho$, its gradient $\nabla\frho$, and the projected density $\prho$ (refer to \cref{sec:closed_form_densities}), we compute the solid geometric constraint (\cref{eq:gs_constr}) in dimensions $d=1,2,3$ and in the limit of infinite resolution:
\begin{align}
\epsilon &= g^s(\lrho,\ell_t)\\
     &= \lim_{N\rightarrow\infty}  \frac{1}{N}\sum_{i=1}^N \bol I_i^s \left[\min\{\frho_i-\eta_e\left(\tfrac{\ell_t}{\Rrad}\right), 0\}\right]^2\\
    &= \frac{1}{\abs{\Omega}}\int_{\Omega} \bol \prho(\nex)e^{-c\abs{\nabla_{\nex}\frho(\nex)}^2} \big[\min\{\frho(\nex) \notag\\ &\qquad\qquad-\eta_e\left(\tfrac{\ell_t}{\Rrad}\right), 0\}\big]^2\de \nex\\
    &=
    \frac{L^{d-1} \Rrad}{\abs{\Omega}}\int_{-\infty}^{\infty} \bol \prho(\tilde x)e^{-\frac{c}{\Rrad^2}\abs{\frac{\mathrm{d}}{\mathrm{d} \tilde x}\frho(\tilde x)}^2} \big[\min\{\frho(\tilde x)\notag\\
    &\qquad\quad-0.75, 0\}\big]^2\de \tilde x  \label{eq:gs_integralform}\\ 
    &= \frac{L^{d-1} \Rrad}{\abs{\Omega}}\left[\frac{3\sqrt{\pi}}{128 \tilde c^{\frac{5}{2}}}\operatorname{erf}\left(\sqrt{\tilde c}\right)-\frac{e^{-\tilde c}}{64\tilde c}\left(2+\frac{3}{\tilde c}\right)\right]\label{eq:epsilon_conic_exact}\\
    &\approx \frac{L^{d-1} \Rrad}{\abs{\Omega}} \frac{3\sqrt{\pi}}{128 \tilde c^{\frac{5}{2}}}, \label{eq:epsilon_apprx}
\end{align}
where the second line corresponds to the integral expression of $g^s$, in the third line we performed the integration along the extra $d-1$ dimensions, yielding a factor of $L^{d-1}$. We also replaced $\eta_e = 0.75$ and changed variables expressing everything in terms of the (scalar) normalized variable $\tilde x = x_1/\Rrad$, which yields the factor of $\Rrad$ in front of the integral and the factor $1/\Rrad^2$ in the exponential. It turns out that $\frho$ and $\prho$ depend on $\tilde x$ rather than on $x$, which simplifies the computation. Moreover, we extended the limits of the integral to $\pm\infty$ since we can always extend $\prho$ to zero outside of the design region. In the fourth line we computed the integral symbolically in terms of the error function $\operatorname{erf}$ and the normalized decay rate $\tilde c = c/\Rrad^2$. Finally, in the last line we approximated $\operatorname{erf}(\sqrt{\tilde c})\approx 1$ and neglected the last term of the previous line, an approximation valid for large~$\tilde c$. This calculation suggests that an appropriate value for the threshold $\epsilon$ is given by \cref{eq:epsilon_apprx}, which depends on the filter radius $\Rrad$, the normalized hyperparameter $\tilde c$, and the dimension $d$. Unfortunately, we see that an appropriate value of~$\epsilon$ also depends on the total length $L$ of the features present in the design, which is specific to the problem and could potentially depend on the choice of radius $\Rrad$. This underscores a limitation of the geometric constraints: the analytical choice of hyperparameters is problem-dependent. Nonetheless, we can still select appropriate hyperparameter values that perform well across a broad range of cases with minimal tuning.

We now briefly discuss an appropriate value for~$c$ to then derive a value for~$\epsilon$. Notice that~$c$ appears in the term $\tilde{g}(x)\coloneqq\exp\left(-c\abs{\tfrac{\de}{\de x}\frho(x)}^2\right)$ of the structural function $\bol I^s$ (\cref{eq:is_func}), indicating that $c$ has units of $[\text{length}]^2$. From the closed-form 1D expression of $\frho$ (\cref{sec:closed_form_densities}), we have that $\tilde{g}$ is a Gaussian $\tilde{g}(x)=\exp\left(-4cx^2/\Rrad^4\right)$ for $x\approx 0$. Thus, writing $c = \Rrad^4/w_c^2$ for a new variable $w_c$, we see that $w_c$ has units of $[\text{length}]$ and corresponds to the length of the interval $\{x\in\R|\, \tilde{g}(x) \geq e^{-1}\}$. Therefore, $w_c$ controls the width of the Gaussian term and, consequently, the width of the inflection region determined by $\bol I^s$ (\cref{eq:is_func}). This width cannot be arbitrarily small nor large. In particular, $w_c$ cannot be much smaller than the pixel size $\Delta x$; otherwise the Gaussian would be too narrow to be accurately discretized. On the other hand, it is difficult to impose a precise lengthscale~$\ell_t$ if the inflection region is too broad, as the imposed lengthscale become overly sensitive to~$\epsilon$. Ideally, $w_c$ should be fraction of $\ell_t\approx \Rrad$, which, in turn, is of the order of a few pixels. It is advisable to set~$w_c$ proportional to~$\Rrad$ so that the threshold~$\epsilon$ (\cref{eq:epsilon_apprx}) depends linearly on~$\Rrad$ and remains independent of~$\ell_t$. In our numerical experiments, we found that $w_c=\Rrad/8$ yields good results in most cases, a choice that corresponds to a decay rate of $c=64\Rrad^2$ or $\tilde{c} = 64$. This value ensures a fast transition of the constraint at $g^s=\epsilon$---as shown in \cref{fig:line_density}c---making the constraint robust to slight variations of~$\epsilon$. With this choice, and the reasonable estimates $L/\abs{\Omega}^{\frac{1}{d}}\approx 1$ and $ \Rrad/\abs{\Omega}^{\frac{1}{d}}\approx10^{-2}$, \cref{eq:epsilon_apprx} yields $\epsilon\approx 10^{-8}$ (we mostly care about the order of magnitude of~$\epsilon$; small variations do not affect the constraint, as discussed above).

\subsection{PDE and bi-PDE filters}\label{sec:pde_bipde_filters}

 PDE-based filtering is popular for unstructured meshes; it requires, however, a different set of analytical hyperparameter values. Furthermore, as described below, special difficulties arise due to the singular Green's function of the modified Helmholtz PDE \eqref{eq:pdefilter_pde}. The mathematical details of the hyperparameter derivations are deferred to \cref{sec:closed_form_densities_pde}, proceeding analogously to the conic filter case. The resulting hyperparameters are summarized in \cref{tab:summary}. Compared to the conic filter case, a different threshold function $\eta_e$ must be employed (\cref{eq:eta_e_pdefilter}), $c$ and $\epsilon$ must be diminished, and $\epsilon$ includes a correction factor $\gamma(\ell_t/\Rrad)$ (\cref{eq:gamma_pde}), for which $\gamma(1)\approx 1.03$.  

Originally, we attempted to use the same value of $c$ as the conic filter case ($c=64\Rrad^2$), but it yielded poor results: in numerous instances, some small features were impossible to eliminate. This phenomenon is due to the presence of large singularity-induced spatial derivatives in the filtered density \correction{$\norm{\nabla_{\nex}\frho(\nex)}$}---particularly in the vicinity of small features. A similar phenomenon was also reported in~\citeasnoun{christiansen_creating_2015}. The large spatial derivatives are attributed to the Green's function $G_{\Rrad}$ of the modified Helmholtz PDE (\cref{eq:green_func_pde}) with effective radius $\Rrad$, which is singular and has unbounded derivatives \correction{$\norm{\nabla_{\nex}G_{\Rrad}}$} at the origin, in 2D and higher dimensions. Since the filtered density can be written as $\frho = G_{\Rrad}\ast\lrho$, we see that $\frho$ inherits large derivatives, which are more prominent in the neighborhood of small features. \correction{(Indeed, for a small feature centered at $\nex_0$, we have that the latent field is approximately proportional to a Dirac delta $\lrho\propto \delta_{\nex_0}$. Thus, $\nabla_{\nex}\frho(\nex) \propto \nabla_{\nex}(G_{\Rrad}\ast\delta_{\nex_0})(\nex)=(\left[\nabla_{\nex}G_{\Rrad}\right]\ast\delta_{\nex_0})(\nex)=\nabla_{\nex}G_{\Rrad}(\nex-\nex_0)$, which exhibits large derivatives at $\nex_0$.)} The solid (and) void structural functions $\bol I^s$ of \cref{eq:is_func}, employs spatial derivative information of $\frho$, weighted by the decay rate $c$, to localize the inflection region where lengthscale violations occur. Thus, the presence of large spatial derivatives (or large numerical errors in their computation) in conjunction with a large $c$ value undermines the ability of the geometric constraints to detect small lengthscale violations. A simple workaround is to reduce the $c$ value: we have found that $c=10 \Rrad^2$ works well in our numerical examples. Unfortunately, a disadvantage of having a lower decay rate $c$ is that lengthscales imposed by the geometric constraint are more sensitive to variations in $\epsilon$; therefore, one may require additional fine-tuning of $\epsilon$ to impose the correct lengthscales.

To devise a more robust filtering approach to use with geometric constraints, we propose an alternative PDE filtering method---denoted ``bi-PDE'' filtering. \correction{Instead of applying the PDE filter with radius $\Rrad$, we simply apply the PDE filter with radius $r\Rrad$ twice in succession, where $r$ is a scaling factor that we adjust to make the effective radius of the bi-PDE filter similar to $\Rrad$ (see details in \cref{sec:closed_form_densities_bipde} and the chosen value of $r$ in \cref{eq:r_0}). With this, the filtered field is now $\frho=(\lrho\ast G_{r\Rrad})\ast G_{r\Rrad}$, or equivalently, $\frho=\lrho\ast (G_{r\Rrad}\ast G_{r\Rrad})$, where $G_{r\Rrad}\ast G_{r\Rrad}$ is the bi-PDE Green's function (\cref{eq:green_func_bipde}), which is well-behaved, has bounded derivatives and is free of singularity}. Consequently, the filtered field does not exhibit large spatial derivatives, rendering the bi-PDE filter more robust under geometric constraints. In contrast to the bi-PDE filter, previous double-filtering techniques~\citep{christiansen_creating_2015,clausen_topology_2015} perform a nonlinear projection in between the two filters, making their effect more difficult to analyze. (Computationally, the cost of topology optimization is usually dominated by the physical PDEs, not the filtering step. Moreover, if explicit sparse-matrix factors are employed, the same factorization of the modified Helmoltz PDE can simply be applied twice.)

We provide in \cref{sec:closed_form_densities_bipde} the details of the bi-PDE filter and the derivation of adequate hyperparameters. The resulting hyperparameters (see \cref{tab:summary}) are similar to the conic filter case, but $\eta_e$ is different (\cref{eq:eta_e_bipdefilter}) and $\epsilon $ includes a correction factor $\gamma(\ell_t/\Rrad)$ (\cref{eq:gamma_bipde}), for which $\gamma(1)\approx 0.973$. 

\section{Algorithm summary}\label{sec:strategy}
In this section, we present our strategy to impose minimum lengthscales on a TO design by means of geometric constraints, in conjunction with SSP binarization and the analytically justified hyperparameter values from \cref{sec:param_deriv}. 

Before starting the optimization, it is important to scale the objective function and constraints (e.g., using affine transformations) to an adequate range, which depends on the optimizer being used. Indeed, many optimization algorithms contain internal ``dimensionful'' parameters---such as initial step-sizes or trust-region radii or penalty strengths---that implicitly assume that all quantities are scaled to have magnitudes of order unity; thus, suboptimal performance is obtained for functions or parameters with inappropriate scalings.  For instance, with the CCSA algorithm \citep{svanberg_class_2002}, Svanberg~\citep{svanberg_mma_2007} suggests scaling the objective function to the range $[1,100]$ and inequality constraints $g(x)-c\leq 0$ such that $c\in [1,100]$. For Ipopt---another gradient-based optimizer for constrained large-scale optimization---the effect of scaling has also been studied~\citep{wachter2006implementation}; by default, it employs a conservative scaling method, in which the objective function and constraints are scaled at the starting point such that the norm of their gradient (in the infinity norm) do not exceed 100. We employ the recommended objective function scalings for the CCSAQ and Ipopt optimizers in all of our examples of \cref{sec:results}. Hereafter, references to the objective function will refer to the \emph{unscaled} version, assuming that the scaling is handled internally. On the other hand, instead of enforcing the geometric constraints as $g^s(\lrho,\ell_t) - \epsilon\leq0$, we normalize them as $\tfrac{g^s(\lrho,\ell_t)}{\epsilon} - 1\leq0$. We explored other constraint variations describing the same feasible set; however, none provided any advantage over the simple $1/\epsilon$ scaling (see Supplementary Information). 

Our strategy to enforce a target minimum lengthscale $\ell_t$ on a TO design proceeds as follows. We have two main stages:
\begin{enumerate}
    \item Unconstrained optimization: we set the filter radius to $\Rrad=\ell_t$ (other options $\Rrad\approx\ell_t$ are also viable, see discussion at the end of \cref{sec:geo_constr}) and optimize a starting design (random, in our case) \emph{without} imposing geometric constraints. We employ SSP binarization, which allows us to rapidly increase the projection steepness $\beta$ across epochs, up to $\beta=\infty$ in the last epoch. As is typically the case in TO, the user is responsible for choosing an appropriate number of iterations, epochs, and $\beta$-scheduling such that the resulting \emph{unconstrained} design achieves a \emph{sufficiently good} objective function value (OFV). 

\item Constrained optimization: we maintain $\beta=\infty$ and impose geometric constraints with the suggested hyperparameter values given in \cref{tab:summary}. The resulting \emph{constrained} design is optimized until a user-defined stopping criterion is met, which usually involves achieving both constraint feasibility \emph{and} an acceptable OFV.
\end{enumerate}

We have found that, upon successful termination, this strategy typically yields a constrained design that achieves a good OFV and complies with the minimum lengthscales imposed by the user, in a small number of iterations---usually converging in fewer iterations than the initial unconstrained optimization---and without requiring any hyperparameter tuning. There are two instances, however, where our strategy fails: when (i) the design fails to satisfy some external lengthscale requirement (such as a foundry design rule~\citep{hammond_photonic_2021}) even though the constraints are feasible, or when (ii) the optimization stalls in a local optimum, impeding further progress. Unfortunately, these cases require the user to rely on heuristic tuning of~$\epsilon$ to obtain a satisfactory design. Regarding the first case, as discussed in \cref{sec:introduction} and \cref{sec:conclusions}, it is a known shortcoming that different definitions of lengthscale constraints do not always \emph{precisely} agree---usually the discrepancy is a fraction of the filter radius. To address this, the user can either: impose much larger lengthscales compared to the desired lengthscale from the beginning, or heuristically decrease $\epsilon$ until the desired lengthscales are imposed, as done in our example of \cref{sec:mode_converter}. Regarding the second case, we found that temporarily imposing a higher threshold~$\epsilon$ for a few iterations may help the optimizer escape local minima---say, imposing $\epsilon=10^{-6}$ for 100 iterations and then $\epsilon=10^{-8}$ until our stopping criteria is met. This approach, however, should be applied judiciously: the constraints are driven out of the feasible region each time~$\epsilon$ is modified, requiring additional iterations to restore the constraints back to feasibility; consequently, repeated updates to~$\epsilon$ could significantly increase the total number of iterations required for convergence.

\section{Examples}\label{sec:results}
Here, we apply our strategy devised in \cref{sec:strategy} to five different TO inverse-design challenges---four from photonics and one from the heat-transfer literature. In particular, we consider three photonics inverse problems introduced in \citeasnoun{schubert_inverse_2022}---the design of a mode converter (\cref{sec:mode_converter}), a beam splitter (\cref{sec:beam_splitter}), and a wavelength demultiplexer (\cref{sec:wdm})---which we simulate using the open-source 2D finite differences frequency domain solver (FDFD) \texttt{Ceviche} \citep{hughes2019forward} via the open-source \texttt{invrs\_gym} package~\citep{schubert2024invrsgymtoolkitnanophotonicinverse}. In addition, we consider the problem of optimizing the resonance of a photonic cavity via the local density of states (LDOS) \citep{Liang:13,icsiklar2022trade}; specifically, we apply our strategy to the 2D cavity design problem defined in \citeasnoun{chen_validation_2024} (\cref{sec:cavity}), which we simulate using \texttt{Ceviche}. Our last example deals with the optimization of the thermal conductivity tensor of a 2D periodic structure (\cref{sec:heat_transfer}), as studied in \citeasnoun{romano_inverse_2022}, which we simulate using an in-house implementation of the finite-volume method (FVM)~\citep{romano2025diffchip}. In each example, the methodology is as follows:\begin{enumerate}
    \item
Unconstrained optimization: we initialize the starting design as random, with the same starting design used for \emph{all} the examples within the same TO challenge. To optimize the unconstrained design we employ the quadratic approximate variation of the CCSA algorithm (CCSAQ) \citep{svanberg_class_2002}, implemented in the open-source library \texttt{NLopt}~\citep{johnson2014nlopt}. The particular $\beta$-scheduling used in each example is given in \cref{sec:unconstr_opt}.
\item
Constrained optimization: we impose geometric constraints. To study if the choice of optimizer affects the performance of the geometric constraints, we employ both CCSAQ and Ipopt~\citep{wachter2006implementation} as optimizers. Each optimizer yields an independent constrained design that have the same unconstrained design as starting point. 

\item Evaluation: We assess the performance of the geometric constraints on the final constrained designs, according to three evaluation metrics described below. The first two measure how far is the constrained design from meeting the desired lengthscales. The third metric is the number of iterations required for termination, with fewer iterations indicating faster convergence.
\end{enumerate}
The evaluation metrics are as follows. First, we consider the solid and void minimum lengthscales---measured using the open-source package \texttt{imageruler}~\citep{chen_validation_2024}---which are then compared against the imposed minimum lengthscale. Ideally, the measured lengthscales should be larger than or equal to the target lengthscale, though \emph{exact} agreement is unlikely because \texttt{imageruler} employs a different definition of ``lengthscale'' based on (non-differentiable) morphological filters. Second, to quantitatively assess how far is a binary design $\prho$ from satisfying the imposed lengthscale constraints---in terms of how many pixels constitute a lengthscale violation---we introduce the \emph{solid violation metric} $\mathcal{V}_s=100\%\cdot v_s/t$, where $v_s$ is number of lengthscale-violating solid pixels and $t$ total number of pixels in $\prho$. The \emph{void violation metric} $\mathcal{V}_v$ is defined analogously. We resort to the package \texttt{imageruler} to determine which solid (void) pixels constitute a lengthscale violation. Lastly, as a third metric we count the number of iterations, in the constrained optimization stage, required to achieve constraints feasibility \emph{and} to attain an OFV sufficiently close to that of the unconstrained design; an efficient constraint would require a small number of iterations to achieve these conditions. Hence, we define our stopping criteria of the second stage as follows: letting $f_u$ and $f_c$ be the OFV of the unconstrained and constrained designs, respectively, which are assumed to be minimized, we optimize the constrained design until one of the following is satisfied: (i) both solid and void constraints are feasible \emph{and} $f_c/f_u\leq 1.25$, or (ii) a defined maximum number of iterations $\mathcal{M}$ ($\mathcal{M}=400$, in our case) is reached. The value 1.25 for the objective function ratio $f_c/f_u$ (OFR) ensures that the performance has not been significantly degraded due to the introduction of the geometric constraints. In some challenging problems (such as the cavity example of \cref{sec:cavity} \citep{chen_validation_2024}), where the objective function is extremely sensitive to both the minimum feature size and small changes in geometry, achieving an OFR close to or less than 1.0 may require numerous iterations---or may even be impossible---when sufficiently large minimum lengthscales are enforced. 

For each TO example we consider three target minimum lengthscales: small, medium (twice the small value), and large (twice the medium value); the small value is specific to each example. In some examples, using the default hyperparameters causes the optimization to become trapped in a local minimum. To address this, we gradually decreased~$\epsilon$ to help the optimizer escape these local minima, as discussed in \cref{sec:strategy}; the affected examples are explicitly identified.

Importantly, \correction{we aggregate outer and inner iterations for both CCSAQ and Ipopt when reporting iterations. This explains why objective function histories may appear irregular and non-monotonic: in an inner iteration, the optimizer may take an aggressive step and subsequently backtrack by increasing some penalty}. In Ipopt, the maximum iteration limit only affects the outer iterations; hence, some of the Ipopt examples may exceed the nominal maximum iteration count.

\subsection{Mode converter}\label{sec:mode_converter}

\begin{figure*}[h!]
\centering
\includegraphics[width=0.7\textwidth]{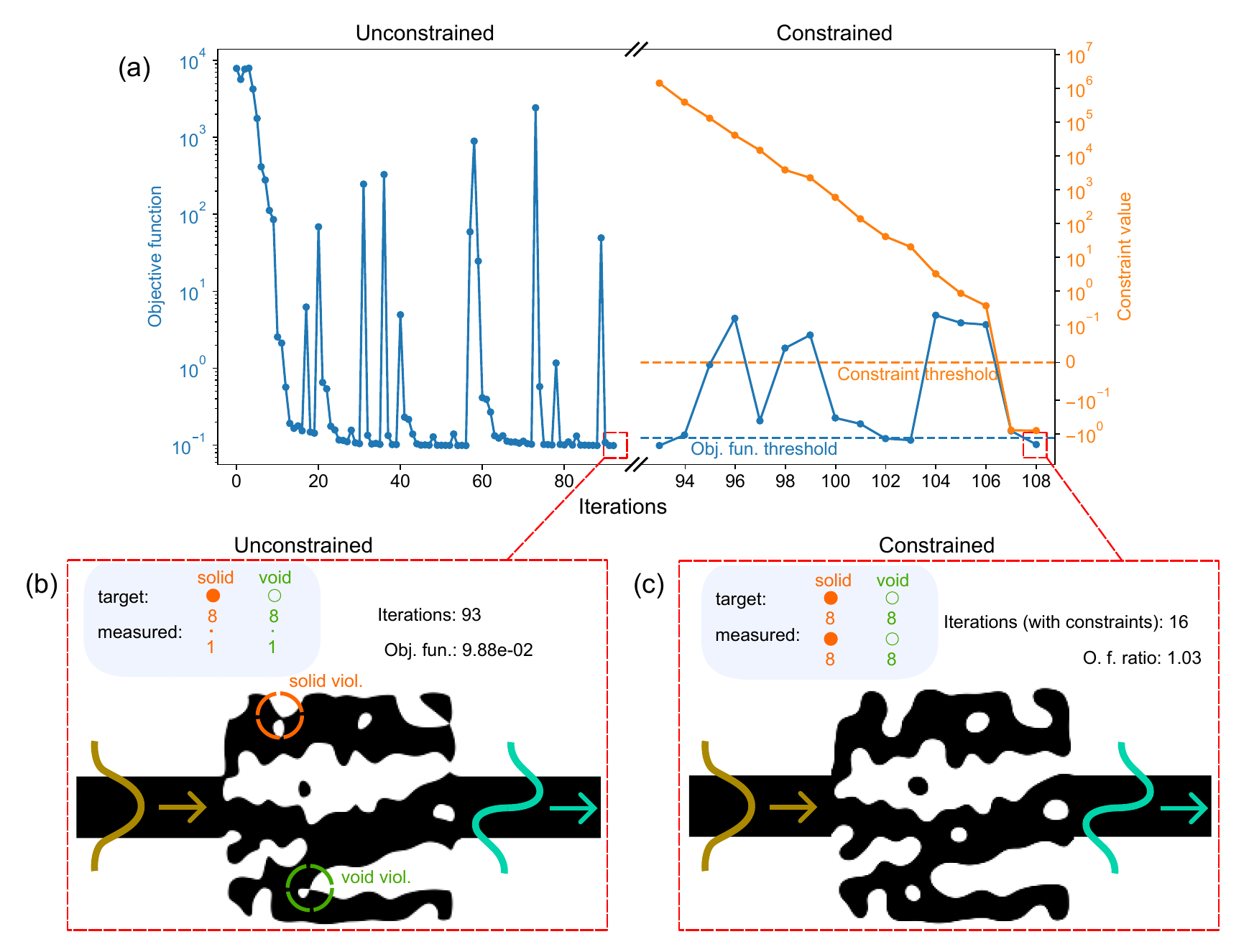} 
\caption{Results for the mode converter example ($\Delta x=$ 10 nm) with medium target lengthscales ($\ell_t= 8\,\Delta x$), using the conic filter (with default hyperparameters $\Rrad=\ell_t$, $c=64\Rrad^2$, and $\epsilon=10^{-8}$), and the CCSAQ optimizer. (a) Objective function history and constraint history (maximum of solid and void constraint) of both the unconstrained and constrained design; the horizontal dashed lines indicate the objective function and constraint thresholds of our stopping criteria. (b) Resulting unconstrained design. (c) Resulting constrained design. In (b) and (c) we indicate the value (in pixels) of the target and measured minimum lengthscales, for both solid and void, which we illustrate with circles whose diameters align with the respective lengthscale. The orange and green dashed circles on the design indicate solid and void lengthscale violations, respectively.} 
\label{fig:mode_converter_result}
\end{figure*}

Following \cite{schubert_inverse_2022}, we consider the problem of designing a 2D photonic mode-converter device that maximally converts the fundamental waveguide mode coming on the left (port 1) to the second order mode on the right (port 2), see \cref{fig:mode_converter_result}b. The design region is a $1.6\times 1.6 \text{ }\mu\text{m}^2$ square discretized with $160\times160$ pixels (pixel size $\Delta x= 10 \text{ nm}$) coupled to fixed and semi-infinite input and output waveguides to the left and right, respectively, having $400 \text{ nm}$ width. The solid and void materials correspond to silicon ($\epsilon_\text{max}=12.25$) and silicon oxide ($\epsilon_\text{min}=2.25$), respectively. Given the incoming fundamental mode from port 1, the goal is to minimize backreflections into port 1 and maximize transmission of the second order mode into port 2, for six wavelengths: 1265, 1270, 1275, 1285, 1290, and 1295 nm. The precise definition of the objective function, which averages backreflection and transmission across all wavelengths, is given in \citeasnoun{schubert_inverse_2022}.

Our results employing the conic filter are shown in \cref{tab:mode_converter}. We observe that all small (4 pixels) and medium (8 pixels) cases achieve a satisfactory constrained design, in less than 21 iterations in all cases, except in the medium/Ipopt case, which requires 66 iterations. We show the objective function and constraint history of the medium/CCSAQ case in \cref{fig:mode_converter_result}, showing that the constraint value decays exponentially with the number of iterations. For the large (16 pixels) cases, the obtained lengthscales fall short of the desired lengthscale by at least 3 pixels; however, the violation metrics indicate that the number of lengthscale-violating pixels is actually small ($\mathcal{V}_s,\mathcal{V}_v < 0.06\%$). We conducted an additional experiment in which we continue evolving the constrained large/CCSAQ case with a decreases threshold of $\epsilon=10^{-12}$: the optimization took 15 extra iterations to meet our stopping criteria, and the resulting constrained design satisfied the minimum lengthscales. We had to decrease~$\epsilon$ by four orders of magnitude because our large $c$-parameter ($\tilde{c} = 64$) renders the constraints relatively insensitive to values of~$\epsilon$ within the same order of magnitude.

Results using the PDE filter are shown in \cref{tab:mode_converter_pde}. Most constrained designs do not satisfy the imposed minimum lengthscales---at most, they are four pixels short. However, the violation metrics indicate that the number of lengthscale-violating pixels is relatively small ($\mathcal{V}_s\leq 0.063\%$, $\mathcal{V}_v\leq 0.16\%$). The loss of precision of the geometric constraints could be attributable to the large spatial derivatives problems caused by singularities in the PDE filter (see \cref{sec:pde_bipde_filters,sec:closed_form_densities_pde}). Even tuning the $\epsilon$ hyperparameter did not cure all violations. The number of iterations in all cases is small---less than 30---similar to the conic filter results.

Results using the bi-PDE filter are shown in \cref{tab:mode_converter_bipde}. Here, most constrained designs comply with the imposed lengthscales---except for the large target lengthscale (16 pixels) case, but even in that case the violation metrics are small ($\mathcal{V}_s=0\%$, $\mathcal{V}_v= 0.016\%$). The number of iterations remains small---less than 40.
\begin{tablefull}[h!]
\centering
\caption{Results using the conic filter for the mode converter example, using default hyperparameters $\Rrad=\ell_t$, $c=64\Rrad^2$, and $\epsilon=10^{-8}$, where $\ell_t$ is the target lengthscale in physical units. Measured lengthscales that comply (do not comply) with the target lengthscale are indicated in green (red). The unconstrained and constrained designs presented in \cref{fig:mode_converter_result} are marked in bold for reference.}
\begin{tabular}{cccccccc}
\toprule
\multicolumn{1}{l|}{}              & \multicolumn{1}{c|}{Algorithm}                               & \multicolumn{2}{c|}{Lengthscales (pix.)}                          & \multicolumn{2}{c|}{Violations (\%)}                   & \multicolumn{2}{c}{Results}                 \\ \hline
\multicolumn{1}{c|}{Target length} & \multicolumn{1}{c|}{Optimizer}  & Solid                           & \multicolumn{1}{c|}{Void}       & $\mathcal{V}_s$ & \multicolumn{1}{c|}{$\mathcal{V}_v$} & Iter.                & Obj. fun. ratio      \\ \midrule
Small: 4 pix.                      &                                                            &                                 &                                 &                 &                                      &                      &                      \\
Unconstrained:                     & CCSAQ                                                     & \color[HTML]{FE0000} 1          & \color[HTML]{FE0000} 2          & 0.13            & 0.054                                & 100                  & 1 \objfun{9.87e-2}          \\ \ignore{\cline{2-9}}
Constrained:                       & CCSAQ                                                 & \color[HTML]{009901} 4          & \color[HTML]{009901} 5          & 0               & 0                                    & 21                   & 1.14                 \\
\multicolumn{1}{l}{}               & IPOPT\footnotemark[1]                                                 & \color[HTML]{009901} 4          & \color[HTML]{009901} 6          & 0               & 0                                    & 19                   & 1.02                 \\ \hline
Medium: 8 pix.                     &\multicolumn{1}{l}{}         & \multicolumn{1}{l}{}            & \multicolumn{1}{l}{}            &                 &                                      & \multicolumn{1}{l}{} & \multicolumn{1}{l}{} \\
Unconstrained:                     & \textbf{CCSAQ}                            & \textbf{\color[HTML]{FE0000} 1}          & \textbf{\color[HTML]{FE0000} 1}          & \textbf{0.21}            & \textbf{0.49}                                 & \textbf{93}                   & \textbf{1 \objfunn{9.88e-2}}   \\ \ignore{\cline{2-9}} 
Constrained:                       & \textbf{CCSAQ}                &\textbf{\color[HTML]{009901} 8} & \textbf{\color[HTML]{009901} 8} & \textbf{0}      & \textbf{0}                           & \textbf{15}          & \textbf{1.03}        \\
                                   & IPOPT                                                 & \color[HTML]{009901} 10         & \color[HTML]{009901} 9          & 0               & 0                                    & 66                   & 0.99     \\ \hline
Large: 16 pix.                     &                                                            &                                 &                                 &                 &                                      &                      &                      \\
Unconstrained:                     & CCSAQ                                                     & \color[HTML]{FE0000} 6          & \color[HTML]{FE0000} 5          & 1.38            & 1.16                                 & 74                   & 1 (1.01e-1)          \\ \ignore{\cline{2-9}} 
Constrained:                       & CCSAQ                                                 & \color[HTML]{FE0000} 15         & \color[HTML]{FE0000} 13         & 0.039           & 0.059                                & 16                   & 1.19                 \\
\multicolumn{1}{l}{}               & IPOPT                                                 & \color[HTML]{FE0000} 15         & \color[HTML]{009901} 17         & 0.0078          & 0                                    & 29                   & 1.03                 \\  \botrule
\end{tabular}
\footnotetext[1]{Used $\epsilon = 10^{-6}$.}
\label{tab:mode_converter}
\end{tablefull}

\begin{tablefull}[h!]
\caption{Results using the PDE filter for the mode converter example, using default hyperparameters $\Rrad=\ell_t$, $c=64\Rrad^2$, and $\epsilon=\left(\gamma(1)\right)^{-3}\cdot10^{-6}$, where $\ell_t$ is the target lengthscale in physical units. Measured lengthscales that comply (do not comply) with the target lengthscale are indicated in green (red).}
\begin{tabular}{cccccccc}
\toprule
\multicolumn{1}{l|}{}              & \multicolumn{1}{c|}{Algorithm}                               & \multicolumn{2}{c|}{Lengthscales (pix.)}           & \multicolumn{2}{c|}{Violations (\%)}                   & \multicolumn{2}{c}{Results}                 \\ \hline
\multicolumn{1}{c|}{Target length} & \multicolumn{1}{c|}{Optimizer}  & Solid                  & \multicolumn{1}{c|}{Void} & $\mathcal{V}_s$ & \multicolumn{1}{c|}{$\mathcal{V}_v$} & Iter.                & Obj. fun. ratio      \\ \midrule
Small: 4 pix.                      &                                                           &                        &                           &                 &                                      &                      &                      \\
Unconstrained:                     & CCSAQ                                                     & \color[HTML]{FE0000}1  & \color[HTML]{FE0000}3     & 0.039           & 0.0078                               & 80                   & 1 \objfun{9.86e-2}          \\ \ignore{\cline{2-9}} 
Constrained:                       & CCSAQ                                                 & \color[HTML]{FE0000}2  & \color[HTML]{FE0000}2     & 0.0078          & 0.0078                               & 15                   & 1.06  \\ \hline
Medium: 8 pix.                     &                               \multicolumn{1}{l}{}         & \multicolumn{1}{l}{}   & \multicolumn{1}{l}{}      &                 &                                      & \multicolumn{1}{l}{} & \multicolumn{1}{l}{} \\
Unconstrained:                     & CCSAQ                                                     & \color[HTML]{FE0000} 4 & \color[HTML]{FE0000} 5    & 0.53            & 0.62                                 & 90                   & 1 \objfunn{9.86e-2}         \\ \ignore{\cline{2-9}} 
Constrained:                       & CCSAQ                                                 & \color[HTML]{FE0000} 7 & \color[HTML]{009901} 10   & 0.0039          & 0                                    & 29                   & 1.19                 \\ \hline
Large: 16 pix.                     &                               &                                                     &                           &                 &                                      &                      &                      \\
Unconstrained:                     & CCSAQ                                                     & \color[HTML]{FE0000} 3 & \color[HTML]{FE0000}7     & 3.28            & 4.07                                 & 85                   & 1 \objfunn{9.95e-2}         \\ \ignore{\cline{2-9}} 
Constrained:                       & CCSAQ\footnotemark[1]                                                 & \color[HTML]{FE0000}12 & \color[HTML]{FE0000}12    & 0.063           & 0.16                                 & 28                   & 1.20                 \\  \botrule              
\end{tabular}
\footnotetext[1]{Used $\epsilon=\left(\gamma(1)\right)^{-3}\cdot10^{-5}$.}
\label{tab:mode_converter_pde}
\end{tablefull}

\begin{tablefull}[h!]
\caption{Results using the Bi-PDE filter for the mode converter example, using default hyperparameters $\Rrad=\ell_t$, $c=64\Rrad^2$, and $\epsilon=\left(\gamma(1)\right)^{-3}\cdot10^{-8}$. Measured lengthscales that comply (do not comply) with the target lengthscale are indicated in green (red).}
\begin{tabular}{cccccccc}
\toprule
\multicolumn{1}{l|}{}              & \multicolumn{1}{c|}{Algorithm}                               & \multicolumn{2}{c|}{Lengthscales (pix.)}           & \multicolumn{2}{c|}{Violations (\%)}                   & \multicolumn{2}{c}{Results}                 \\ \hline
\multicolumn{1}{c|}{Target length} & \multicolumn{1}{c|}{Optimizer}  & Solid                  & \multicolumn{1}{c|}{Void} & $\mathcal{V}_s$ & \multicolumn{1}{c|}{$\mathcal{V}_v$} & Iter.                & Obj. fun. ratio      \\ \midrule
Small: 4 pix.                      &                               &                                                     &                           &                 &                                      &                      &                      \\
Unconstrained:                     & CCSAQ                                                     & \color[HTML]{FE0000}3  & \color[HTML]{009901}5     & 0.0078          & 0                                    & 92                   & 1 \objfun{9.85e-2}   \\ \ignore{\cline{2-9}} 
Constrained:                       & CCSAQ                                                 & \color[HTML]{009901}6  & \color[HTML]{009901}7     & 0               & 0                                    & 17                   & 1.07  \\ \hline
Medium: 8 pix.                     &                               \multicolumn{1}{l}{}         & \multicolumn{1}{l}{}   & \multicolumn{1}{l}{}      &                 &                                      & \multicolumn{1}{l}{} & \multicolumn{1}{l}{} \\
Unconstrained:                     & CCSAQ                                                     & \color[HTML]{FE0000}4  & \color[HTML]{FE0000} 3    & 0.10            & 0.20                                 & 80                   & 1 \objfunn{9.84e-2}   \\ \ignore{\cline{2-9}} 
Constrained:                       & CCSAQ                                                 & \color[HTML]{009901}8  & \color[HTML]{009901} 9    & 0               & 0                                    & 17                   & 1.11                 \\ \hline
Large: 16 pix.                     &                               &                                                      &                           &                 &                                      &                      &                      \\
Unconstrained:                     & CCSAQ                                                     & \color[HTML]{FE0000}1  & \color[HTML]{FE0000}3     & 1.38            & 2.34                                 & 160                  & 1 \objfunn{1.62e-1}   \\ \ignore{\cline{2-9}} 
Constrained:                       & CCSAQ\footnotemark[1]                                                 & \color[HTML]{009901}17 & \color[HTML]{FE0000}13    & 0               & 0.016                                & 38                   & 1.00                 \\ \botrule              
\end{tabular}
\footnotetext[1]{Used $\epsilon= \left(\gamma(1)\right)^{-3}\cdot10^{-9}$.}
\label{tab:mode_converter_bipde}
\end{tablefull}

\subsection{Beam splitter}\label{sec:beam_splitter}

\begin{figure*}[h!]
\centering
\includegraphics[width=0.7\textwidth]{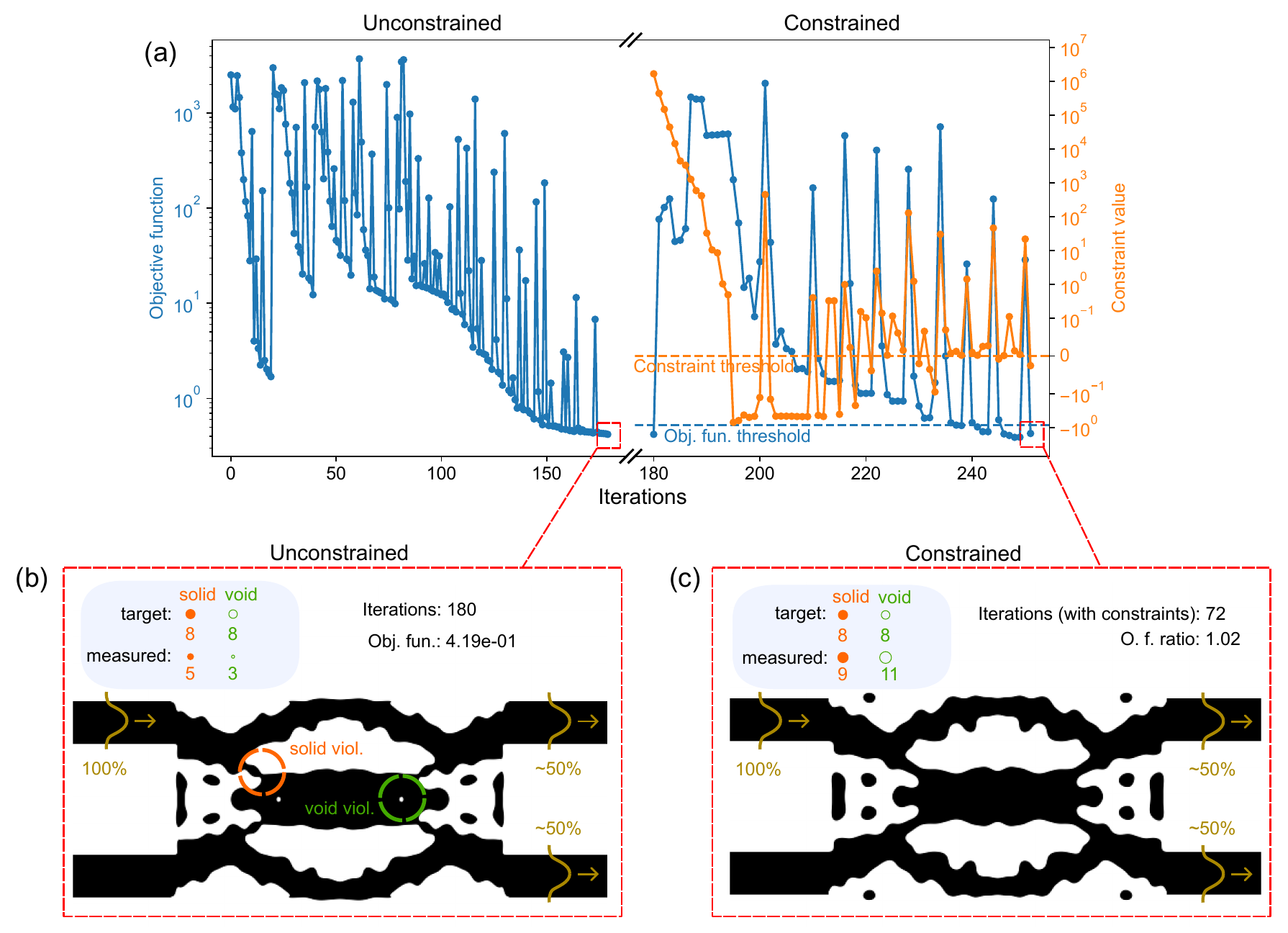} 
\caption{Results for the beam splitter example ($\Delta x=$ 10 nm) with medium target lengthscales ($\ell_t= 8\,\Delta x$), using the conic filter (with default hyperparameters $\Rrad=\ell_t$, $c=64\Rrad^2$, and $\epsilon=10^{-8}$), and the CCSAQ optimizer. (a) Objective function history and constraint history (maximum of solid and void constraint) of both the unconstrained and constrained design; the horizontal dashed lines indicate the objective function and constraint thresholds of our stopping criteria. (b) Resulting unconstrained design. (c) Resulting constrained design. In (b) and (c) we indicate the value (in pixels) of the target and measured minimum lengthscales, for both solid and void, which we illustrate with circles whose diameters align with the respective lengthscale. The orange and green dashed circles on the design indicate solid and void lengthscale violations, respectively.} 
\label{fig:beam_splitter}
\end{figure*}

We consider the design of the 2D photonics beam splitter defined in \cite{schubert_inverse_2022}. The device has four ports, ports 1 and 3 to the left and ports 2 and 3 to the right, connected to fixed and semi-infinite waveguides of $400 \text{ nm}$ width, see \cref{fig:beam_splitter}b. The design region, at the center, is a $3.2\times 2.0 \text{ }\mu\text{m}^2$ rectangle discretized with $320\times200$ pixels (pixel size $\Delta x= 10 \text{ nm}$). An incoming fundamental mode into port 1 is divided equally in power into ports 2 and 3, while minimizing backreflection into port 1 and transmission into port 4. Symmetry along the vertical and horizontal axis is imposed so that all ports have analogous behavior. The solid and void material permittivities, and the six considered wavelengths, are the same as the mode converter example of \cref{sec:mode_converter}.

The results obtained with the conic filter are shown in \cref{tab:beam_splitter}. The small (4 pixels) cases comply with the imposed lengthscales, but the CCSAQ and Ipopt optimizers exhibit a large difference of the required number of iterations---39 and 288, respectively. For the medium (8 pixels) cases, all constrained designs comply with the imposed lengthscale in less than 72 iterations. In the objective function and constraint history of the medium/CCSAQ case (\cref{fig:beam_splitter}), we note that the geometric constraints become feasible exponentially fast---in less than 20 iterations---but then, the CCSAQ optimizer requires $\sim50$ more iterations to recover the performance lost while driving the constraints to feasibility. The large cases (16 pixels), on the other hand, do not comply with the target lengthscale---they fall short by five pixels, at most---whereas the violation metrics are small, but not negligible ($\mathcal{V}_s \sim 0.38\%$). The iteration count is considerable: it takes 374 iterations at most to meet the stopping criteria. This suggests that the beam splitter becomes a challenging example when geometric constraints with large lengthscales are imposed. We conducted an additional experiment in which we continue optimizing the large/CCSAQ constrained design with $\epsilon=10^{-13}$: the resulting design has 16 and 15 pixels of solid and void lengthscales, respectively, with $\mathcal{V}_s=0\%$ and $\mathcal{V}_v=0.00625\%$, and it took 185 additional iterations to reach our stopping criteria.

\begin{figure*}[h!]
\centering
\includegraphics[width=0.7\textwidth]{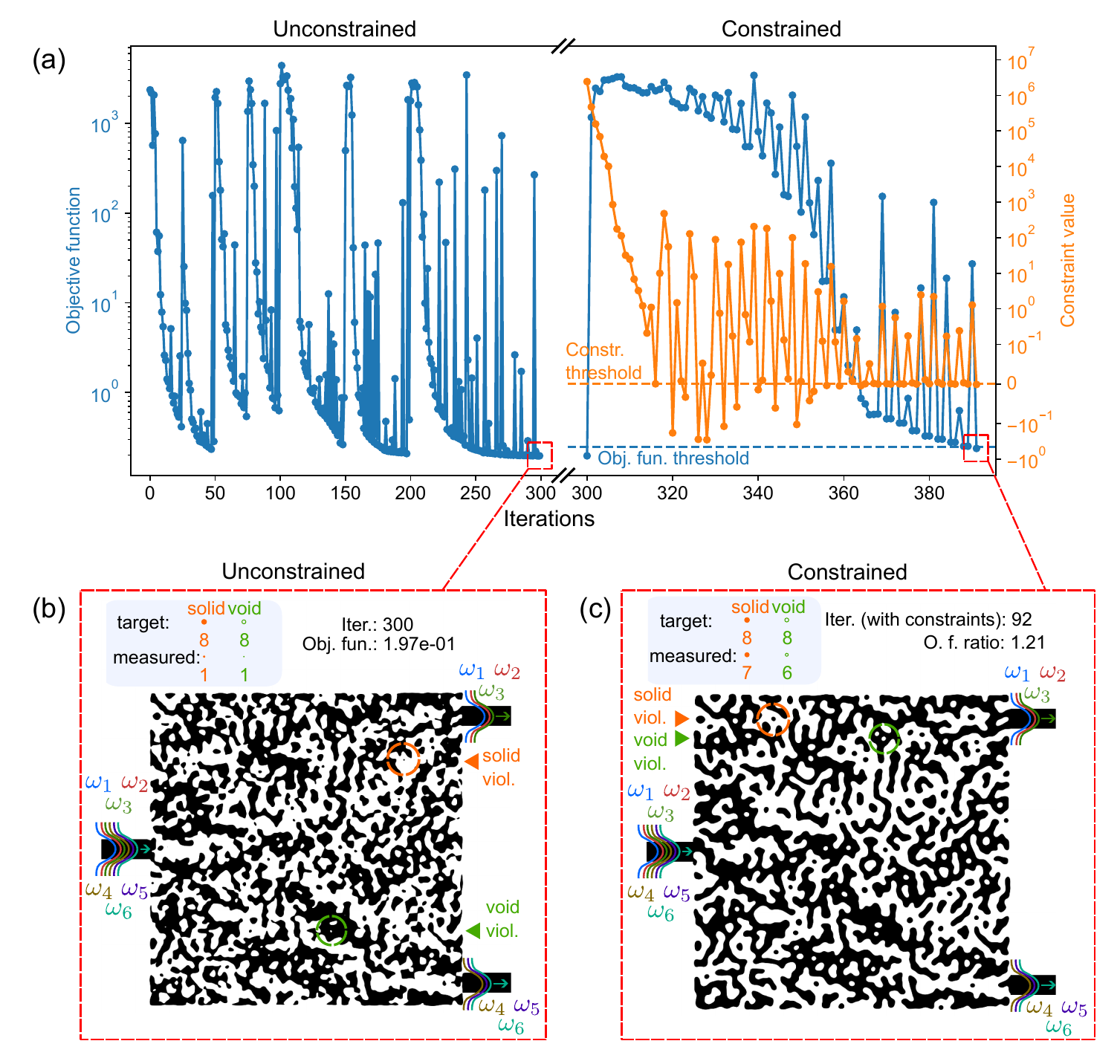} 
\caption{Results for the wavelength demultiplexer example ($\Delta x=$ 10 nm) with medium target lengthscales ($\ell_t= 8\,\Delta x$), using the conic filter (with default hyperparameters $\Rrad=\ell_t$, $c=64\Rrad^2$, and $\epsilon=10^{-8}$), and the CCSAQ optimizer. (a) Objective function history and constraint history (maximum of solid and void constraint) of both the unconstrained and constrained design; the horizontal dashed lines indicate the objective function and constraint thresholds of our stopping criteria. (b) Resulting unconstrained design. (c) Resulting constrained design. In (b) and (c) we indicate the value (in pixels) of the target and measured minimum lengthscales, for both solid and void, which we illustrate with circles whose diameters align with the respective lengthscale. The orange and green dashed circles on the design indicate solid and void lengthscale violations, respectively.} 
\label{fig:wdm}
\end{figure*}

\begin{tablefull}[h!]
\caption{Results using the conic filter for the beam splitter example, using default hyperparameters $\Rrad=\ell_t$, $c=64\Rrad^2$, and $\epsilon=10^{-8}$, where $\ell_t$ is the target lengthscale in physical units. Measured lengthscales that comply (do not comply) with the target lengthscale are indicated in green (red). The unconstrained and constrained designs presented in \cref{fig:beam_splitter} are marked in bold for reference.}
\begin{tabular}{ccccccccc}
\toprule
\multicolumn{1}{l|}{}              & \multicolumn{1}{c|}{Algorithm}                               & \multicolumn{2}{c|}{Lengthscales (pix.)}                            & \multicolumn{2}{c|}{Violations (\%)}                        & \multicolumn{2}{c}{Results}                 \\ \hline
\multicolumn{1}{c|}{Target length} & \multicolumn{1}{c|}{Optimizer} & Solid                            & \multicolumn{1}{c|}{Void}        & $\mathcal{V}_s$ & \multicolumn{1}{c|}{$\mathcal{V}_v$} & Iter.                & Obj. fun. ratio      \\ \midrule
Small: 4 pix.                      &                               &                              &                                  &                                  &                 &                                      &                      &                      \\
Unconstrained:                     & CCSAQ                                                     & \color[HTML]{FE0000} 1           & \color[HTML]{FE0000} 2           & 0.18            & 0.038                                & 180                  & 1 \objfun{2.50e-1}          \\ \ignore{\cline{2-9}} 
Constrained:                       & {CCSAQ}                               & {\color[HTML]{009901} 4}  & {\color[HTML]{009901} 5}  & {0}      & {0}                           & {39}          & {1.19}        \\
\multicolumn{1}{l}{}               & IPOPT                                                 & \color[HTML]{009901} 4           & \color[HTML]{009901} 5           & 0               & 0                                    & 288                  & 0.8                  \\ \hline
Medium: 8 pix.                     &                                \multicolumn{1}{l}{}         & \multicolumn{1}{l}{}             & \multicolumn{1}{l}{}             &                 &                                      & \multicolumn{1}{l}{} & \multicolumn{1}{l}{} \\
Unconstrained:                     & \textbf{CCSAQ}                                                     & \textbf{\color[HTML]{FE0000} 5}           & \textbf{\color[HTML]{FE0000} 3}           & \textbf{0.33}            & \textbf{0.0125}                               & \textbf{180}                  & \textbf{1} \textbf{\objfunn{4.19e-1}}   \\ \ignore{\cline{2-9}} 
Constrained:                       & \textbf{CCSAQ}                             & \textbf{\color[HTML]{009901} 9}  & \textbf{\color[HTML]{009901} 11} & \textbf{0}      & \textbf{0}                           & \textbf{72}          & \textbf{1.02}        \\
                                   & IPOPT                                                 & \color[HTML]{009901} 9           & \color[HTML]{009901} 9           & 0               & 0                                    & 50                   & 0.64    \\ \hline
Large: 16 pix.                     &                               &                                                               &                                  &                 &                                      &                      &                      \\
Unconstrained:                     & CCSAQ                                                    & \color[HTML]{FE0000} 6           & \color[HTML]{FE0000} 1           & 2.43            & 0.63                                 & 200                  & 1 (3.49e0)           \\ \ignore{\cline{2-9}} 
Constrained:                       & {CCSAQ}\footnotemark[1]                              & {\color[HTML]{FE0000} 11} & {\color[HTML]{FE0000} 14} & {0.38}  & {6.3e-3}                      & {374}         & {0.96}        \\
\multicolumn{1}{l}{}               & IPOPT                                                 & \color[HTML]{FE0000} 14          & \color[HTML]{FE0000} 13          & 0.35            & 0.025                                & 320                  & 0.4                  \\  \botrule
\end{tabular}
\footnotetext[1]{Used $\epsilon=10^{-3}$ for 200 iterations, then $\epsilon=10^{-8}$ for 174 iterations.}
\label{tab:beam_splitter}
\end{tablefull}

Results using the PDE filter are shown in \cref{tab:beam_splitter_pde}. Both the medium (8 pixels) and small (4 pixels) cases comply with the imposed target lengthscale, taking 76 and 15 iterations, respectively. For the large case (16 pixels), the void lengthscale falls short by four pixels---although $\mathcal{V}_v=0.025\%$ is small. It reached the maximum 400 iterations: the objective function decreased so slowly that it could not reach our stopping criteria, as evidenced in the resulting OFR of $\sim 40$. This slow convergence was also observed in the conic filter results.

\begin{tablefull}[h!]
\caption{Results using the PDE filter for the beam splitter example, using default hyperparameters $\Rrad=\ell_t$, $c=64\Rrad^2$, and $\epsilon=\left(\gamma(1)\right)^{-3}\cdot10^{-6}$. Measured lengthscales that comply (do not comply) with the target lengthscale are indicated in green (red).}
\begin{tabular}{cccccccc}
\toprule
\multicolumn{1}{l|}{}              & \multicolumn{1}{c|}{Algorithm}                               & \multicolumn{2}{c|}{Lengthscales (pix.)}           & \multicolumn{2}{c|}{Violations (\%)}                   & \multicolumn{2}{c}{Results}                 \\ \hline
\multicolumn{1}{c|}{Target length} & \multicolumn{1}{c|}{Optimizer}  & Solid                  & \multicolumn{1}{c|}{Void} & $\mathcal{V}_s$ & \multicolumn{1}{c|}{$\mathcal{V}_v$} & Iter.                & Obj. fun. ratio      \\ \midrule
Small: 4 pix.                      &                               &                                                      &                           &                 &                                      &                      &                      \\
Unconstrained:                     & CCSAQ                                                     & \color[HTML]{FE0000}3  & \color[HTML]{FE0000}1     & 0.025           & 0.088                                & 180                  & 1 \objfun{2.10e-1}  \\ \ignore{\cline{2-9}} 
Constrained:                       & CCSAQ                                                 & \color[HTML]{009901}5  & \color[HTML]{009901}5     & 0               & 0                                    & 15                   & 1.15  \\ \hline
Medium: 8 pix.                     &                                \multicolumn{1}{l}{}         & \multicolumn{1}{l}{}   & \multicolumn{1}{l}{}      &                 &                                      & \multicolumn{1}{l}{} & \multicolumn{1}{l}{} \\
Unconstrained:                     & CCSAQ                                                     & \color[HTML]{FE0000}4  & \color[HTML]{FE0000}3     & 0.11            & 0.075                                & 180                  & 1 \objfunn{2.50e-1}   \\ \ignore{\cline{2-9}} 
Constrained:                       & CCSAQ                                                 & \color[HTML]{009901}9  & \color[HTML]{009901}8     & 0               & 0                                    & 76                   & 1.24                 \\ \hline
Large: 16 pix.                     &                                                             &                        &                           &                 &                                      &                      &                      \\
Unconstrained:                     & CCSAQ                                                     & \color[HTML]{FE0000}3  & \color[HTML]{FE0000}4     & 5.67            & 2.55                                 & 200                  & 1 \objfunn{1.23e0}   \\ \ignore{\cline{2-9}} 
Constrained:                       & CCSAQ\footnotemark[1]                                                 & \color[HTML]{009901}20 & \color[HTML]{FE0000}12    & 0               & 0.025                                & 400                  & 41.94                \\  \botrule              
\end{tabular}
\footnotetext[1]{Reached maximum number of iterations. Objection function is converging slowly.} 
\label{tab:beam_splitter_pde}
\end{tablefull}

Results using the bi-PDE filter are shown in \cref{tab:beam_splitter_bipde}. We observe similar results to those of the PDE filter. Notably, the medium and small cases take a few more iterations---133 and 57, respectively. Furthermore, for the large case the solid violation is much higher ($\mathcal{V}_s=0.24\%$); however, by the end of the 400 maximum iterations, it achieved a lower OFR of 3.24.

\begin{tablefull}[h!]
\caption{Results using the Bi-PDE filter for the beam splitter example, using default hyperparameters $\Rrad=\ell_t$, $c=64\Rrad^2$, and $\epsilon=\left(\gamma(1)\right)^{-3}\cdot10^{-8}$. Measured lengthscales that comply (do not comply) with the target lengthscale are indicated in green (red).}
\begin{tabular}{cccccccc}
\toprule
\multicolumn{1}{l|}{}              & \multicolumn{1}{c|}{Algorithm}                               & \multicolumn{2}{c|}{Lengthscales (pix.)}           & \multicolumn{2}{c|}{Violations (\%)}                   & \multicolumn{2}{c}{Results}                 \\ \hline
\multicolumn{1}{c|}{Target length} & \multicolumn{1}{c|}{Optimizer} & Solid                  & \multicolumn{1}{c|}{Void} & $\mathcal{V}_s$ & \multicolumn{1}{c|}{$\mathcal{V}_v$} & Iter.                & Obj. fun. ratio      \\ \midrule
Small: 4 pix.                      &                               &                                                      &                           &                 &                                      &                      &                      \\
Unconstrained:                     & CCSAQ                                                     & \color[HTML]{FE0000}2  & \color[HTML]{FE0000}2     & 0.075           & 0.0125                               & 180                  & 1 \objfun{0.21e-1}  \\ \ignore{\cline{2-9}} 
Constrained:                       & CCSAQ                                                 & \color[HTML]{009901}5  & \color[HTML]{009901}5     & 0               & 0                                    & 57                   & 1.25   \\ \hline
Medium: 8 pix.                     &                                \multicolumn{1}{l}{}         & \multicolumn{1}{l}{}   & \multicolumn{1}{l}{}      &                 &                                      & \multicolumn{1}{l}{} & \multicolumn{1}{l}{} \\
Unconstrained:                     & CCSAQ                                                     & \color[HTML]{FE0000}3  & \color[HTML]{FE0000}2     & 0.18            & 0.16                                 & 180                  & 1 \objfunn{3.21e-1}   \\ \ignore{\cline{2-9}} 
Constrained:                       & CCSAQ\footnotemark[1]                                                 & \color[HTML]{009901}9  & \color[HTML]{FE0000}7     & 0               & 0.044                                & 133                  & 1.04                 \\ \hline
Large: 16 pix.                     &                                                             &                        &                           &                 &                                      &                      &                      \\
Unconstrained:                     & CCSAQ                                                     & \color[HTML]{FE0000}7  & \color[HTML]{FE0000}4     & 2.56            & 3.05                                 & 419                  & 1 \objfunn{1.61e0}   \\ \ignore{\cline{2-9}} 
Constrained:                       & CCSAQ\footnotemark[2]                                                 & \color[HTML]{FE0000}11 & \color[HTML]{FE0000}15    & 0.24            & 0.094                                & 400                  & 3.24                 \\ \botrule             
\end{tabular}
\footnotetext[1]{Used $\epsilon=\left(\gamma(1)\right)^{-3}\cdot10^{-9}$.}
\footnotetext[2]{Reached maximum number of iterations. Objection function is converging slowly.}
\label{tab:beam_splitter_bipde}
\end{tablefull}

\subsection{Wavelength demultiplexer}\label{sec:wdm}

In this example we consider the design of the 2D photonic wavelength demultiplexer defined in~\citeasnoun{schubert_inverse_2022}. The design region is a $6.4\times 6.4 \text{ }\mu\text{m}^2$ square discretized with $640\times 640$ pixels (pixel size $\Delta x= 10 \text{ nm}$). The device is coupled to an input waveguide to the left (port 1) and two output waveguides to the right (port 2 and 3), of $400 \text{ nm}$ width, see \cref{fig:wdm}b. Incoming excitations from port 1 are directed to port 2 for wavelengths 1265, 1270, and 1275 nm, and directed to port 3 for wavelengths 1285, 1290, and 1295 nm. The solid and void materials correspond to silicon ($\epsilon_\text{max}=12.25$) and silicon oxide ($\epsilon_\text{min}=2.25$), respectively.

The results using the conic filter are shown in \cref{fig:wdm,tab:wdm}. All small (4 pixels) cases satisfy the imposed lengthscale. We note that CCSAQ required significantly fewer iterations than Ipopt---66 and 262, respectively---which was also observed in the beam-splitter example. The medium (8 pixels) cases require less than 124 iterations, but they fall short of the imposed lengthscale by at most two pixel. The violation metrics, however, show that the number of violating pixels is small ($\mathcal{V}_s,\mathcal{V}_d < 0.02\%$). Similar to the beam-splitter example, we observe in the constraint history (\cref{fig:wdm}a) that the constraints become feasible in less 20 iterations, but then CCSAQ takes an additional $\sim 70$ iterations to meet our stopping criteria. For the large (16 pixels) cases, all final designs fail to comply with the desired lengthscale by at most four pixels; however, the violation metrics are small ($\mathcal{V}_s,\mathcal{V}_d < 0.04\%$), and our stopping criteria was met in a small number of iterations---less than 80. It is surprising that the wavelength demultiplexer requires significantly fewer iterations than the beam splitter (\cref{sec:beam_splitter}) for the same target lengthscale, despite its higher resolution and seemingly greater design complexity with many small features. We conducted an additional experiment in which we continue optimizing the large/CCSAQ constrained design with $\epsilon=10^{-15}$: the resulting design required 94 additional iterations to satisfy our stopping criteria and meets the imposed lengthscale of 16 pixels.

\begin{tablefull}[h!]
\caption{Results using the conic filter for the wavelength demultiplexer example, using default hyperparameters $\Rrad=\ell_t$, $c=64\Rrad^2$, and $\epsilon=10^{-8}$, where $\ell_t$ is the target lengthscale in physical units. Measured lengthscales that comply (do not comply) with the target lengthscale are indicated in green (red). The unconstrained and constrained designs presented in \cref{fig:wdm} are marked in bold for reference.}
\begin{tabular}{cccccccc}
\toprule
\multicolumn{1}{l|}{}              & \multicolumn{1}{c|}{Algorithm}                               & \multicolumn{2}{c|}{Lengthscales (pix.)}                          & \multicolumn{2}{c|}{Violations (\%)}                   & \multicolumn{2}{c}{Results}                 \\ \hline
\multicolumn{1}{c|}{Target length} & \multicolumn{1}{c|}{Optimizer}  & Solid                           & \multicolumn{1}{c|}{Void}       & $\mathcal{V}_s$ & \multicolumn{1}{c|}{$\mathcal{V}_v$} & Iter.                & Obj. fun. ratio      \\ \midrule
Small: 4 pix.                      &                               &                                                               &                                 &                 &                                      &                      &                      \\
Unconstrained:                     & CCSAQ                                                     & \color[HTML]{FE0000} 1          & \color[HTML]{FE0000} 1          & 0.11            & 0.059                                & 267                  & 1 \objfun{1.97e-1}          \\ \ignore{\cline{2-9}} 
Constrained:                       & CCSAQ                                                 & \color[HTML]{009901} 4          & \color[HTML]{009901} 4          & 0               & 0                                    & 66                   & 1.20                 \\
\multicolumn{1}{l}{}               & IPOPT                                                 & \color[HTML]{009901} 4          & \color[HTML]{009901} 4          & 0               & 0                                    & 262                  & 1.01                 \\ \hline
Medium: 8 pix.                     &                                \multicolumn{1}{l}{}         & \multicolumn{1}{l}{}            & \multicolumn{1}{l}{}            &                 &                                      & \multicolumn{1}{l}{} & \multicolumn{1}{l}{} \\
Unconstrained:                     & \textbf{CCSAQ}                                                    & \textbf{\color[HTML]{FE0000} 1}          & \textbf{\color[HTML]{FE0000} 1}          & \textbf{1.11}            & \textbf{1.0}                                  & \textbf{300}                  & \textbf{1 \objfunn{1.97e-1}}   \\ \ignore{\cline{2-9}} 
Constrained:                       & \textbf{CCSAQ}                        & \textbf{\color[HTML]{FE0000} 7} & \textbf{\color[HTML]{FE0000} 6} & \textbf{0.013}  & \textbf{0.016}                       & \textbf{92}          & \textbf{1.21}        \\
                                   & IPOPT                                                & \color[HTML]{009901} 8          & \color[HTML]{FE0000} 7          & 0               & 0.00049                              & 124                  & 1.06    \\ \hline
Large: 16 pix.                     &                               &                                                               &                                 &                 &                                      &                      &                      \\
Unconstrained:                     & CCSAQ                                                     & \color[HTML]{FE0000} 2          & \color[HTML]{FE0000} 2          & 1.93            & 1.13                                 & 300                  & 1 (2.35e-1)          \\ \ignore{\cline{2-9}} 
Constrained:                       & CCSAQ                                                 & \color[HTML]{FE0000} 11         & \color[HTML]{FE0000} 13         & 0.037           & 0.009                                & 78                   & 1.24                 \\
\multicolumn{1}{l}{}               & IPOPT                                                 & \color[HTML]{FE0000} 14         & \color[HTML]{FE0000} 15         & 0.013           & 0.016                                & 56                   & 0.88                 \\  \botrule
\end{tabular}
\label{tab:wdm}
\end{tablefull}

\subsection{Cavity design}\label{sec:cavity}

\begin{figure*}[h!]
\centering
\includegraphics[width=0.7\textwidth]{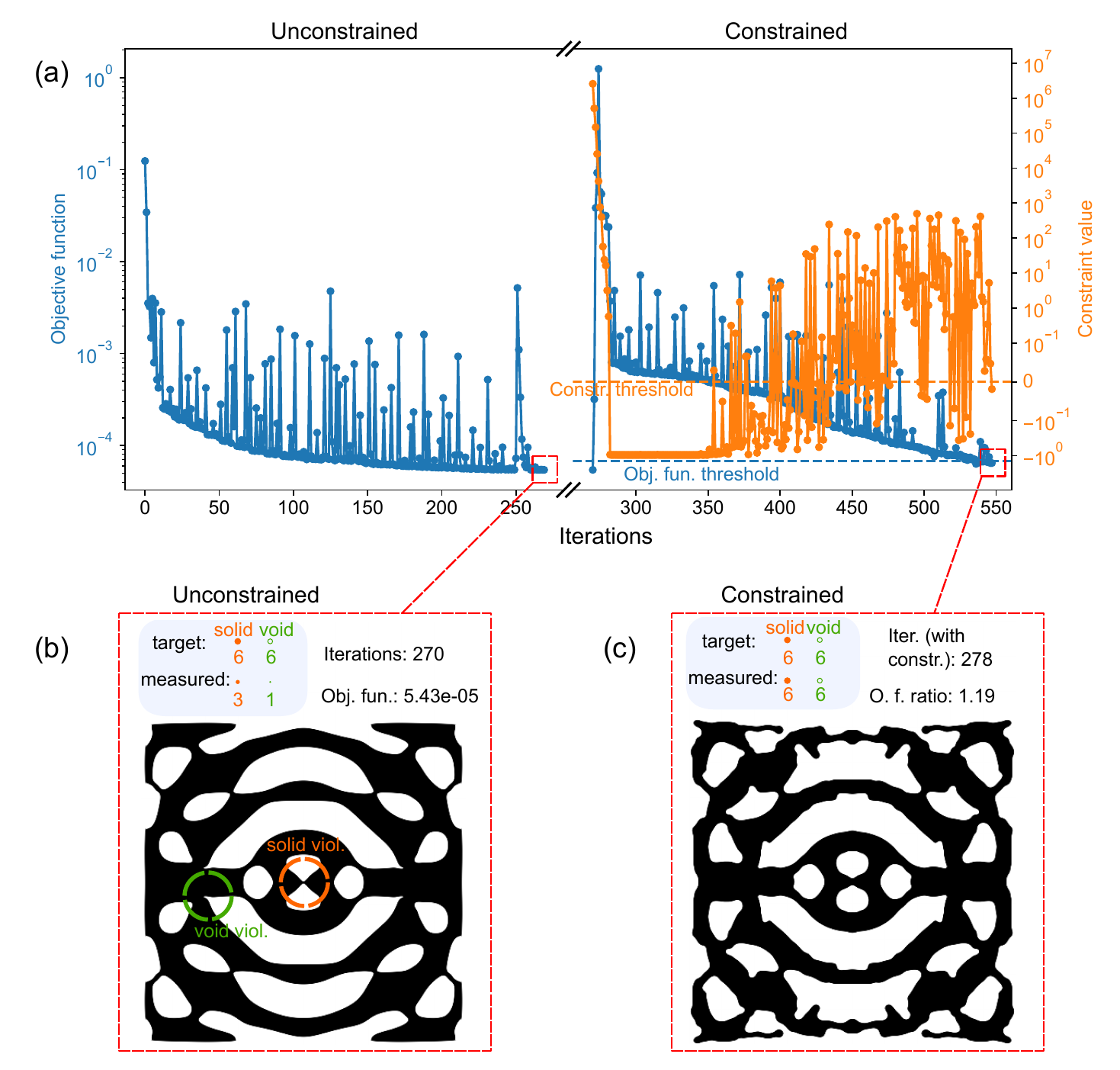} 
\caption{Results for the cavity example ($\Delta x=$ 5 nm) with medium target lengthscales ($\ell_t= 6\,\Delta x$), using the conic filter (with default hyperparameters $\Rrad=\ell_t$, $c=64\Rrad^2$, and $\epsilon=10^{-8}$), and the CCSAQ optimizer. (a) Objective function history and constraint history (maximum of solid and void constraint) of both the unconstrained and constrained design; the horizontal dashed lines indicate the objective function and constraint thresholds of our stopping criteria. (b) Resulting unconstrained design. (c) Resulting constrained design. In (b) and (c) we indicate the value (in pixels) of the target and measured minimum lengthscales, for both solid and void, which we illustrate with circles whose diameters align with the respective lengthscale. The orange and green dashed circles on the design indicate solid and void lengthscale violations, respectively.} 
\label{fig:cavity}
\end{figure*}

In this example, we consider the maximization of the local density of states (LDOS) of a 2D photonic cavity, corresponding to the power emitted by a point-dipole source. This TO problem is challenging due to its extreme sensitivity to the imposed minimum lengthscale and small geometric changes~\citep{chen_validation_2024}---especially for resonances with long lifetimes, corresponding to a high ``quality factor'' $Q$~\citep{joannopoulos2008photonic}. As a result, the optimization is ill-conditioned and ``stiff'', causing many algorithms to converge slowly~\citep{Liang:13}. This challenge is thus useful for testing the geometric constraints performance. 

In particular, we consider the cavity design problem defined in~\citeasnoun{chen_validation_2024}. The goal is to maximize the LDOS produced by an unit electric dipole with in-plane electric field polarization, for a wavelength of $\lambda=1.55\text{ }\mu\text{m}$. The dipole is placed at the center of the design region, consisting of a $2.005\times2.005\text{ }\mu\text{m}^2$ square discretized with $401\times 401$ pixels (pixel size $\Delta x= 5 \text{ nm}$). The solid and void materials have permittivities $\epsilon_\text{max}=12.11$ and $\epsilon_\text{min}=1$, respectively. The cavity  is surrounded by void material with outgoing boundary conditions, implemented via perfectly matched layers (PML)~\citep{taflove_em_book}. As done in~\citeasnoun{chen_validation_2024}, instead of directly maximizing the cavity LDOS, we equivalently minimize the non-dimensionalized objective function:
\begin{align}
    f = \frac{\text{vacuum LDOS}}{\text{cavity LDOS}},
\end{align}
where the cavity LDOS depends on the material distribution determined by $\lrho$, whereas the vacuum LDOS is numerically computed with the same discretization and $\lrho=0$ everywhere. This objective function is more suitable for optimizing high-$Q$ cavities, resulting in faster convergence~\citep{Liang:13}. Due to the inherent slow convergence of this example, we increased the maximum number of iterations for the constrained optimization to $\mathcal{M}=800$.

In \cref{fig:cavity}, we show the objective function and constraint history of the medium/CCSAQ (6 pixels) case, using a conic filter. The slow convergence of the cavity LDOS optimization is evident during the unconstrained optimization: In the first $\sim15$ iterations the objective function achieves a value of $\sim2\times10^{-4}$, but then it takes $\sim250$ additional iterations to decrease by one order of magnitude, to a value of $\sim5\times10^{-5}$. When geometric constraints are imposed and the constrained optimization begins, the constraints achieve feasibility exponentially fast in only $\sim 20$ iteration; however, this fast decrease of the constraint comes at the expense of degrading the objective function, which increased by one order of magnitude to a value of $\sim5\times10^{-4}$. The CCSAQ optimizer then required $\sim250$ additional iterations to decrease the objective function back to its unconstrained value and meet our stopping criteria. Although the constrained optimization required approximately the same number of iterations as the unconstrained case, we note that the inclusion of geometric constraints did not affect the convergence rate of the objective function, which remained similar to that of the unconstrained optimization.

\begin{figure*}[h!]
\centering
\includegraphics[width=0.7\textwidth]{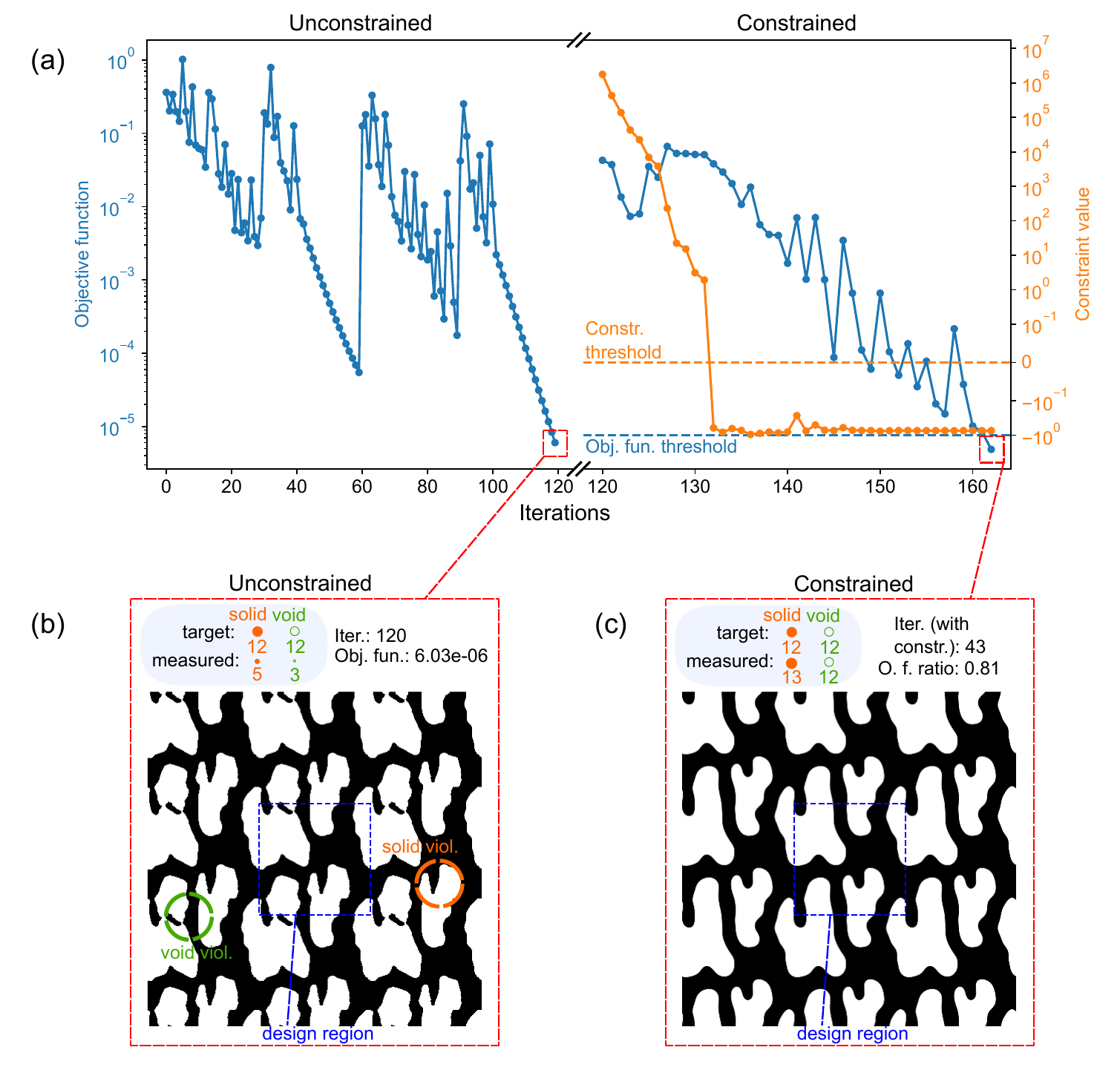} 
\caption{Results for the heat-transfer example with medium target lengthscales ($\ell_t= 6\,\Delta x$), using the conic filter (with default hyperparameters $\Rrad=\ell_t$, $c=64\Rrad^2$, and $\epsilon=10^{-8}$), and the CCSAQ optimizer. (a) Objective function history and constraint history (maximum of solid and void constraint) of both the unconstrained and constrained design; the horizontal dashed lines indicate the objective function and constraint thresholds of our stopping criteria. (b) Resulting unconstrained design. (c) Resulting constrained design. In (b) and (c) we indicate the value (in pixels) of the target and measured minimum lengthscales, for both solid and void, which we illustrate with circles whose diameters align with the respective lengthscale. The orange and green dashed circles on the design indicate solid and void lengthscale violations, respectively.} 
\label{fig:heat_transfer}
\end{figure*}

Our results using the conic filter are shown in \cref{tab:cavity}. For the small (3 pixels) target lengthscale, both the CCSAQ and Ipopt cases comply with the imposed lengthscale, but they require vastly different numbers of iterations---21 and 1745, respectively. For the medium (6 pixels) target, the CCSAQ case satisfies the imposed lengthscale, requiring 287 iterations to meet our stopping criteria. On the other hand, the Ipopt case reaches the maximum number of iterations without satisfying the geometric constraints and falls below the imposed solid target by 3 pixels; however, the solid violation is small ($\mathcal{V}_s=0.0044\%$). For the large (12 pixels) target, the CCSAQ case requires 717 iterations but exhibits a small void lengthscale violation ($\mathcal{V}_v=0.017\%$), whereas the Ipopt case complies with the imposed lengthscale but reaches the maximum number of iterations without meeting our stopping criteria.

\begin{tablefull}[h!]
\caption{Results using the conic filter for the cavity example, using default hyperparameters $\Rrad=\ell_t$, $c=64\Rrad^2$, and $\epsilon=10^{-8}$, where $\ell_t$ is the target lengthscale in physical units. Measured lengthscales that comply (do not comply) with the target lengthscale are indicated in green (red). The unconstrained and constrained designs presented in \cref{fig:wdm} are marked in bold for reference.}
\begin{tabular}{cccccccc}
\toprule
\multicolumn{1}{l|}{}              & \multicolumn{1}{c|}{Algorithm}                               & \multicolumn{2}{c|}{Lengthscales (pix.)}                          & \multicolumn{2}{c|}{Violations (\%)}                   & \multicolumn{2}{c}{Results}                 \\ \hline
\multicolumn{1}{c|}{Target length} & \multicolumn{1}{c|}{Optimizer}  & Solid                           & \multicolumn{1}{c|}{Void}       & $\mathcal{V}_s$ & \multicolumn{1}{c|}{$\mathcal{V}_v$} & Iter.                & Obj. fun. ratio      \\ \midrule
Small: 3 pix.                      &                                                             &                                 &                                 &                 &                                      &                      &                      \\
Unconstrained:                     & CCSAQ                                                   & \color[HTML]{FE0000} 2          & \color[HTML]{009901} 3          & 0.01            & 0                                    & 270                  & 1 \objfun{5.12e-5}          \\ \ignore{\cline{2-9}} 
Constrained:                       & CCSAQ                                                 & \color[HTML]{009901} 3          & \color[HTML]{009901} 4          & 0               & 0                                    & 21                   & 0.65                 \\
\multicolumn{1}{l}{}               & IPOPT\footnotemark[1]                                                 & \color[HTML]{009901} 4          & \color[HTML]{009901} 3          & 0               & 0                                    & 1745                  & 0.76                 \\ \hline
Medium: 6 pix.                     &                                \multicolumn{1}{l}{}         & \multicolumn{1}{l}{}            & \multicolumn{1}{l}{}            &                 &                                      & \multicolumn{1}{l}{} & \multicolumn{1}{l}{} \\
Unconstrained:                     & \textbf{CCSAQ}                                          & \textbf{\color[HTML]{FE0000} 3}          & \textbf{\color[HTML]{FE0000} 1}          & \textbf{0.11}            & \textbf{0.03}                                 & \textbf{270}                  & \textbf{1 \objfunn{5.43e-5}}   \\ \ignore{\cline{2-9}} 
Constrained:                       & \textbf{CCSAQ}                 & \textbf{\color[HTML]{009901} 6} & \textbf{\color[HTML]{009901} 6} & \textbf{0}      & \textbf{0}                           & \textbf{287}         & \textbf{1.19}        \\
                                   & IPOPT\footnotemark[2]                                                 & \color[HTML]{FE0000} 3          & \color[HTML]{009901} 7          & 0.0044
               & 0                                & 1261                  & 0.53      \\ \hline
Large: 12 pix.                     &                               &                                                               &                                 &                 &                                      &                      &                      \\
Unconstrained:                     & CCSAQ                                                    & \color[HTML]{FE0000} 3          & \color[HTML]{FE0000} 10         & 0.008           & 0.021                                & 270                  & 1 (8.03e-5)          \\ \ignore{\cline{2-9}} 
Constrained:                       & CCSAQ                                                 & \color[HTML]{009901} 13         & \color[HTML]{FE0000} 10         & 0          & 0.017                               & 717                  & 1.23                 \\
\multicolumn{1}{l}{}               & IPOPT                                                 & \color[HTML]{009901} 15         & \color[HTML]{009901} 14         & 0               & 0                                    & 1338                  & 0.49                 \\ \botrule
\end{tabular}
\footnotetext[1]{Reached maximum number of iterations. Constraints do not become feasible.}
\footnotetext[2]{Reached maximum number of iterations.}
\label{tab:cavity}
\end{tablefull}

\subsection{Heat transfer}\label{sec:heat_transfer}

In this section, we apply our framework to optimize a 2D periodic thermal metamaterial with a prescribed effective thermal conductivity tensor, $\bar{\kappa}_D$. The transport model is the standard heat-conduction equation
\begin{equation}\label{eq:heat}
\nabla \cdot  \left[\kappa(\boldsymbol \rho)\nabla  T  \right]=0 , 
\end{equation}
which we solve using the finite-volume method~\citep{gersborg2006topology}. The design region is a square of size $L$ divided into $150\times 150$ pixels, see \cref{fig:heat_transfer}b. Details about the discretization of \cref{eq:heat} can be found in Ref.~\cite{romano_inverse_2022}. The thermal metamaterial is composed of a solid and a void phase, the latter with negligible thermal conductivity. In \cref{eq:heat}, the local thermal conductivity is interpolated via
\begin{equation}
\kappa(\prho) = \delta  + \left(\kappa_{\text{bulk}}-\delta\right)\prho,
\end{equation}
where $\kappa_{\text{bulk}}$ = 1 Wm$^{-1}$K$^{-1}$ is the bulk thermal conductivity of the solid phase and $\delta = 10^{-10}$ Wm$^{-1}$K$^{-1}$ is a regularization parameter. The goal is to find $\lrho$ such that the {effective thermal conductivity tensor} $\bar{\kappa}(\prho)$ approaches the desired one, which we set to
\begin{equation}
\bar{\kappa}_D = \begin{bmatrix} 0.2 & 0 \\
0 & 0.3
\end{bmatrix} \text{ Wm$^{-1}$K$^{-1}$}.
\end{equation}
To reconstruct $\bar{\kappa}(\prho)$, we solve \cref{eq:heat} for three different directions of applied temperature $\Delta T_\text{ext}\, \mathbf{\hat{n}}_i$, with $\Delta T_\text{ext}=1$ K, $\mathbf{\hat{n}}_1 = \mathbf{\hat{x}}$, $\mathbf{\hat{n}}_2 = \mathbf{\hat{y}}$, and $\mathbf{\hat{n}}_3 = \tfrac{\sqrt{2}}{2}\left(\mathbf{\hat{x}} +\mathbf{\hat{y}} \right)$. For each direction $i=1,2,3$, we compute the effective \emph{directional} thermal conductivity $\kappa_i =\Delta T_\text{ext}^{-1}\mathbf{P}_i\cdot\mathbf{\hat{n}}_i$, where $\mathbf{P}_i = [P_{(i,x)},P_{(i,y)}]$ is the vector of power collected along two orthogonal faces of the simulation domain. For example, the component along $x$ is computed by
\begin{equation}
P_{(i,x)} = \int_{-L/2}^{L/2} \mathbf{J}_i(L/2,y)\cdot \hat{\mathbf{x}} \de y,
\end{equation}
where $\mathbf{J} = -\kappa \nabla T$ is the heat flux.~%Note that we use arbitrary values for $\Delta T$ and $L$ since, for a given $\prho$, their values do not alter $\bar{\kappa}$.
The three independent components $\bar{\kappa}_{xx}$, $\bar{\kappa}_{yy}$, and $\bar{\kappa}_{xy}$ of the symmetric $2\times2$ tensor $\bar{\kappa}$ are then obtained by solving the linear system $\kappa_i = \mathbf{\hat{n}}_i^\top  \bar{\kappa}\,\mathbf{\hat{n}}_i = \hat{n}_{i,x}^2\bar{\kappa}_{xx}+\hat{n}_{i,y}^2\kappa_{yy}+2\hat{n}_{i,x}\hat{n}_{i,y}\bar{\kappa}_{xy}$, leading to $\bar{\kappa}_{xx} = \kappa_1$, $\bar{\kappa}_{yy} = \kappa_2$, and $\bar{\kappa}_{xy} = -\tfrac{1}{2}\kappa_1-\tfrac{1}{2}\kappa_2 + \kappa_3$. Once $\bar{\kappa}(\prho)$ is reconstructed, the objective function is computed as $f(\prho) = \norm{\bar{\kappa}(\prho) - \bar{\kappa}_D}$, where we used the Frobenius norm. 

End-to-end differentiability is ensured by automatic differentiation, which is the engine behind the thermal solver. The code has recently been validated against Ansys
on three-dimensional structures \citep{romano2025diffchip}, and is planned for public release in the near future. We note that the use of three different directions generalizes our previous approach \citep{romano_inverse_2022}, where only the diagonal components of $\bar \kappa$ were considered. In contrast, here, the third direction is used to enforce zero off-diagonal elements. In principle, a diagonal tensor for $\bar \kappa$ can also be achieved by enforcing symmetry. However, our approach allows for a wider range of optimal configurations, while also potentially enabling non-zero off-diagonal components. 

Our results using the conic filter are shown in \cref{fig:heat_transfer}. Only two final designs do not satisfy the imposed lengthscales; however, their number of lengthscale-violating pixels is small ($\mathcal{V}_s,\mathcal{V}_v<0.05\%$). All examples meet our stopping conditions in less than 150 iterations. For the small case, Ipopt requires more than triple number of iterations compared to CCSAQ (141 and 43 iterations, respectively). For the medium and large cases, Ipopt only requires a few extra iterations compared to CCSAQ. In \cref{fig:heat_transfer} we present the objective function and constraint history of the medium/CCSAQ case, which exhibit rapid convergence.

\begin{tablefull}[h!]
\caption{Results using the conic filter for the heat-transfer example, using default hyperparameters $\Rrad=\ell_t$, $c=64\Rrad^2$, and $\epsilon=10^{-8}$, where $\ell_t$ is the target lengthscale in physical units. Measured lengthscales that comply (do not comply) with the target lengthscale are indicated in green (red). The unconstrained and constrained designs presented in \cref{fig:heat_transfer} are marked in bold for reference.}
\begin{tabular}{cccccccc}
\toprule
\multicolumn{1}{l|}{}              & \multicolumn{1}{c|}{Algorithm}                               & \multicolumn{2}{c|}{Lengthscales (pix.)}                            & \multicolumn{2}{c|}{Violations (\%)}                   & \multicolumn{2}{c}{Results}                 \\ \hline
\multicolumn{1}{c|}{Target length} & \multicolumn{1}{c|}{Optimizer}  & Solid                            & \multicolumn{1}{c|}{Void}        & $\mathcal{V}_s$ & \multicolumn{1}{c|}{$\mathcal{V}_v$} & Iter.                & Obj. fun. ratio      \\ \midrule
Small: 6 pix.                      &                                                             &                                  &                                  &                 &                                      &                      &                      \\
Unconstrained:                     & CCSAQ                                                     & \color[HTML]{FE0000} 1           & \color[HTML]{FE0000} 1           & 2.14            & 1.13                                 & 120                  & 1 \objfun{2.65e-6}          \\ \ignore{\cline{2-9}} 
Constrained:                       & CCSAQ                                                 & \color[HTML]{FE0000} 5           & \color[HTML]{FE0000} 5           & 0.0044          & 0.018                                & 43                   & 1.17                 \\
\multicolumn{1}{l}{}               & IPOPT                                                 & \color[HTML]{009901} 7           & \color[HTML]{009901} 6           & 0               & 0                                    & 141                  & 0.95                 \\  \hline
Medium: 12 pix.                    &                                \multicolumn{1}{l}{}         & \multicolumn{1}{l}{}             & \multicolumn{1}{l}{}             &                 &                                      & \multicolumn{1}{l}{} & \multicolumn{1}{l}{} \\
Unconstrained:                     & \textbf{CCSAQ}                                                  & \textbf{\color[HTML]{FE0000} 5}           & \textbf{\color[HTML]{FE0000} 3}           & \textbf{1.42}            & \textbf{0.35}                                 & \textbf{120}                  & \textbf{1 \objfunn{6.03e-6}}   \\ \ignore{\cline{2-9}} 
Constrained:                       & \textbf{CCSAQ}                & \textbf{\color[HTML]{009901} 13} & \textbf{\color[HTML]{009901} 12} & \textbf{0}      & \textbf{0}                           & \textbf{43}          & \textbf{0.81}        \\
                                   & IPOPT                                                 & \color[HTML]{009901} 16          & \color[HTML]{009901} 15          & 0               & 0                                    & 67                   & 1.19                 \\ \hline
Large: 18 pix.                     &                                                             &                                  &                                  &                 &                                      &                      &                      \\
Unconstrained:                     & CCSAQ                                                     & \color[HTML]{FE0000} 4           & \color[HTML]{FE0000} 5           & 4.55            & 0.85                                 & 120                  & 1 (5.29e-4)          \\ \ignore{\cline{2-9}} 
Constrained:                       & CCSAQ                                                 & \color[HTML]{009901} 20          & \color[HTML]{FE0000} 14          & 0               & 0.044                                & 67                   & 0.80                 \\
\multicolumn{1}{l}{}               & IPOPT                                                 & \color[HTML]{009901} 21          & \color[HTML]{009901} 23          & 0               & 0                                    & 85                   & 3.8e-3               \\ \botrule
\end{tabular}
\label{tab:heat_transfer}
\end{tablefull}

\section{Discussion}\label{sec:discussion}
Our strategy is able to, either, impose the desired minimum lengthscale $\ell_t$, or yield a design whose lengthscale falls slightly below the $\ell_t$ but with only a few lengthscale-violating pixels---as measured by $\mathcal{V}_s$ and $\mathcal{V}_v$. 

Usually, for small and medium target lengthscales, the geometric constraints successfully enforce the desired lengthscale, in a small number of iterations---typically, less than that of the unconstrained optimization---and without degrading the objective function value. On the other hand, the large target lengthscale cases are more challenging: often, lengthscales are not satisfied and it takes numerous iterations to meet our stopping criteria. A possible reason for this is that, since the perimeter of 2D features are longer for large target lengthscales, there are more pixels that can constitute a possible lengthscale violation, and thus it is more difficult to perfectly agree with lengthscale measuring tools. The resulting violation metrics, however, are small (often $\mathcal{V}_s,\mathcal{V}_v < 0.1\%$), and we have shown that they can be further reduced, potentially to zero, by decreasing~$\epsilon$. A possible cause for the increased number of iterations is that larger lengthscales---which require a larger filter radius $\Rrad$---have a slower convergence rate. Indeed, we observe that, for the same number of iterations, the unconstrained large-lengthscale designs attain a less-optimal objective function compared to unconstrained medium-lengthscale cases. Furthermore, it is usually the case that the optimizer cannot recover the same objective function value as the unconstrained design when sufficiently large lengthscales are imposed, thus leading to an increased number of iterations to meet our stopping conditions. We note that the constraints rapidly become feasible within a few iterations, which initially worsens the objective function value; a large number of subsequent iterations are then needed to restore its performance. Notably, when small or medium target lengthscales are imposed, the geometric constraints do not appear to degrade the convergence rate relative to the unconstrained case, as evidenced in the challenging cavity example shown in \cref{fig:cavity}a. 

Comparing the performance of the optimizers, we find that CCSAQ usually requires fewer iterations than Ipopt to meet our stopping criteria, which can be attributed to differences in their constrained optimization approaches. On one hand, when the starting point is infeasible, the CCSAQ implementation of the \texttt{NLopt} library~\citep{johnson2014nlopt} employs an aggressive strategy that essentially disregards the objective function value and prioritize constraint feasibility instead, potentially leading to fewer iterations but also becoming more susceptible to converging to local minima. On the other hand, Ipopt employs a conservative approach: by default, it allows some constraint infeasibility while prioritizing objective function improvement; this enhances robustness but could lead to a higher iteration count.

Regarding the conic, PDE, and bi-PDE filter cases, they exhibit similar performance. It is important to emphasize, however, that lengthscales imposed by geometric constraints become more sensitive to~$\epsilon$ when the PDE filter is employed---as discussed in \cref{sec:pde_bipde_filters}; therefore, users might expect additional fine-tuning of $\epsilon$ to enforce strict minimum lengthscales. The conic and bi-PDE filters, on the other hand, do not encounter this issue, as their filter kernels are bounded.

\section{Conclusions}\label{sec:conclusions}
In this work we provided a comprehensive analysis of the geometric constraints and the analytical derivation of adequate and resolution-invariant hyperparameters, for the case of full binarization ($\beta
=\infty$) and infinite resolution. We proposed a simple but effective strategy to impose minimum lengthscales on a TO design via geometric \correction{constraints}, based on our analytically derived hyperparameters and the use of subpixel-smooth projection (SSP), which allows the rapidly-converging optimization of almost-binary designs. We illustrated and evaluated the performance of our strategy using four photonics and one heat-transfer TO examples, each constrained by three distinct target lengthscales, while employing different combination of filters (conic, PDE, and bi-PDE) and optimizers (CCSAQ and Ipopt). We believe that our streamlined algorithm will be attractive to adopt in many TO applications. Lastly, we identified that the PDE filter poses difficulties to the geometric constraints due to its singular Green's function. We proposed the bi-PDE filter scheme, which is a simple and attractive alternative that alleviates the aforementioned issues due to its well-behaved Green's function.

There are many directions of future research. For example, one could apply a similar analysis (in the full-binarization and infinite-resolution limit) to other types of fabrication constraints, such as maximum lengthscales \citep{guest_imposing_2009,lazarov_maximum_2017,carstensen_projection_2018}, minimum area and minimum enclosed area \citep{hammond_photonic_2021}, and connectivity \citep{li_structural_2016,christiansen_inverse_2023}. Another challenge is that the hyperparameter $\epsilon$ (\cref{eq:epsilon_apprx}) has a weak dependence on the total volume of the small features due to the integrated nature of the constraint, which sometimes still requires manual tuning of~$\epsilon$. One possible way to eliminate this problem-dependence is to impose the constraints not just globally but locally. Although this would require thousands of additional pointwise constraints, the sparsity of their gradients could be exploited by efficient large-scale optimization algorithms~\citep[e.g.,][]{wachter2006implementation,Gill2005,Byrd2006}. We must emphasize, however, that imposing an even-stricter minimum lengthscale may not necessarily improve the manufacturability of a design, because lengthscale constraints are generally approximations. The definition of a minimum lengthscale is inherently somewhat ambiguous for freeform geometries, especially at the scale of the discretization~\citep{chen_validation_2024}, so 1--2 pixel violations are not necessarily meaningful. More fundamentally, however, minimum lengthscales are generally only a proxy for the true manufacturing constraints of a process like lithography or 3D printing. In practice, additional calibration steps---such as proximity error correction (PEC)---are required to make TO designs manufacturable; nonetheless, these steps often produce output patterns that deviate from the original TO design, leading to reduced performance~\citep{zhou_topology_2014}. One approach to improve fidelity is to incorporate an explicit model of the fabrication physics into the design framework~\citep{zhou_topology_2014,zhou_topology_2017,zheng_manufacturing2023}, which has the drawback of requiring extensive theory and calibration for a particular manufacturing method.

\begin{appendices}
\section{Derivations: conic~filters}\label{sec:closed_form_densities}
Here we derive closed-form expressions of $\frho$ and $\prho$ for the simplified 1D setting presented in \cref{sec:param_deriv}, in the limit of infinite resolution and full binarization ($\beta=\infty$). Consider a latent density $\lrho$ that consists of a 1D binary bump of width $h$, $\lrho(x) = \operatorname{rect}(x/h)$, for $x\in\R$, as done in \cref{eq:rho_strip}. For filtering, we employ the 1D conic filter (\cref{eq:conic_filter}) of radius $\Rrad$ and normalization $a=1/\Rrad$, which can be written as
$$
\filtk_\Rrad(x) = \frac{1}{\Rrad}\,\Lambda\left(\frac{x}{\Rrad}\right),
$$
in terms of the triangular function $\Lambda(x) = \operatorname{max}\{1-\abs{x},0\}$.
It is convenient to work with normalized units $\tilde x = x/\Rrad$ and $\tilde h = h/\Rrad$. In the limit of infinite resolution, the filtered field is:
\begin{align}
\frho(x) &= \left(\filtk_\Rrad*\lrho\right)(x)\\
&= \int_{\R} \filtk_\Rrad(x-t)\lrho(t)\de t \\
&= \int_\R \frac{1}{\Rrad}\,\Lambda\left(\frac{x-t}{R}\right)\operatorname{rect}\left(\frac{t}{h}\right)\de t\\
&= \int_{-\tfrac{h}{2}}^{\tfrac{h}{2}} \frac{1}{\Rrad}\,\Lambda\left(\frac{x-t}{R}\right)\de t\\
&= \int_{-\tfrac{\tilde h}{2}}^{\tfrac{\tilde  h}{2}}\Lambda\left(\tilde{x}-\tilde{t}\right)\de \tilde  t,
\end{align}
where in the last line we changed variables $\tilde t = t/\Rrad$. From the last expression we see that $\frho$ depends on $\tilde x$ and $\tilde{h}$ rather than on $x$ and $h$. Calculating this in closed-form yields a piecewise quadratic polynomial whose expression depends on $\tilde{h}$. For the case $0<\tilde h \leq 1$ we have:
\begin{align}
\frho&(x) = \notag\\
&\begin{cases}
    -\abs{\tilde x}^2+\frac{\tilde h}{4}(4-\tilde h), &\abs{\tilde x}\in\left[0,\frac{\tilde h}{2}\right),\\
    \tilde h(1-\abs{\tilde x}),  &\abs{\tilde x}\in\left[\frac{\tilde h}{2},1-\frac{\tilde h}{2}\right),\\
    \frac{1}{2}\left(1-\left(\abs{\tilde x}-\frac{\tilde h}{2}\right)\right)^2,  &\abs{\tilde x}\in\left[1-\frac{\tilde h}{2},1+\frac{\tilde h}{2}\right),\\
    0,  &\abs{\tilde x}\in\left[1+\frac{\tilde h}{2},\infty\right),
\end{cases} \label{eq:rhotilde_1}
\end{align}
and for the case $1<\tilde h\leq 2$:
\begin{align}
    \frho&(x) = \notag\\ 
    &\begin{cases}
        -\abs{\tilde x}^2+\frac{\tilde h}{4}(4-\tilde h),  &\abs{\tilde x}\in\left[0,1-\frac{\tilde h}{2}\right),\\
        -\frac{1}{8}\left((\tilde h-2(1+\abs{\tilde x}))^2-8\right), &\abs{\tilde x}\in\left[1-\frac{\tilde h}{2},\frac{\tilde h}{2}\right),\\
        \frac{1}{2}\left(1-\left(\abs{\tilde x}-\frac{\tilde h}{2}\right)\right)^2,  &\abs{\tilde x}\in\left[\frac{\tilde h}{2},1+\frac{\tilde h}{2}\right),\\
        0, &\abs{\tilde x}\in\left[1+\frac{\tilde h}{2},\infty\right). \label{eq:rhotilde_2}
    \end{cases}
\end{align}
The case $\tilde h > 2$ is not of interest, since the geometric constraint can only impose lengthscales up to $2\Rrad$ when employing the conic filter, but we show its expression here for completeness: 
\begin{align}
    \frho&(x) = \notag\\ 
    &\begin{cases}
        1,  &\abs{\tilde x}\in\left[0,\frac{\tilde h}{2}-1\right),\\
        -\frac{1}{8}\left((\tilde h-2(1+\abs{\tilde x}))^2-8\right), &\abs{\tilde x}\in\left[\frac{\tilde h}{2}-1,\frac{\tilde h}{2}\right),\\
        \frac{1}{2}\left(1-\left(\abs{\tilde x}-\frac{\tilde h}{2}\right)\right)^2,  &\abs{\tilde x}\in\left[\frac{\tilde h}{2},1+\frac{\tilde h}{2}\right),\\
        0,  &\abs{\tilde x}\in\left[1+\frac{\tilde h}{2},\infty\right). \label{eq:rhotilde_3}
    \end{cases}
\end{align}
Now, the projected density is computed as a Heaviside projection $(\beta=\infty)$, that is $\prho = H(\frho-1/2)$, and it can be shown that it corresponds to a binary bump of width $w$, $\prho(x) = \operatorname{rect}(x/w)$. The (normalized) width $\tilde w = w/\Rrad$ can be computed in closed-form (by solving $\frho(w/2)=1/2$) as a function of the (normalized) latent width $\tilde h$~\citep{qian_topological_2013}:
\begin{align}\label{eq:l_to_h}
    \tilde w \left(\tilde h\right) = 
    \begin{cases}
        0, &\quad \tilde h\in \left[0,2-\sqrt{2}\right),\\
        \sqrt{4\tilde h-\tilde h^2-2}, &\quad \tilde h\in \left[2-\sqrt{2},1\right),\\
        \tilde h, &\quad \tilde h\in [1,\infty),
    \end{cases}
\end{align}
whereas the the inverse relation is:
\begin{align}\label{eq:h_to_l}
    \tilde h \left(\tilde w\right) = 
    \begin{cases}
        2-\sqrt{2-\tilde w^2}, &\quad \tilde w \in (0,1),\\
        \tilde w, &\quad \tilde w\in [1,\infty).        
    \end{cases}
\end{align}

\begin{figure}[h!]
\centering
\includegraphics[width=0.4\textwidth]{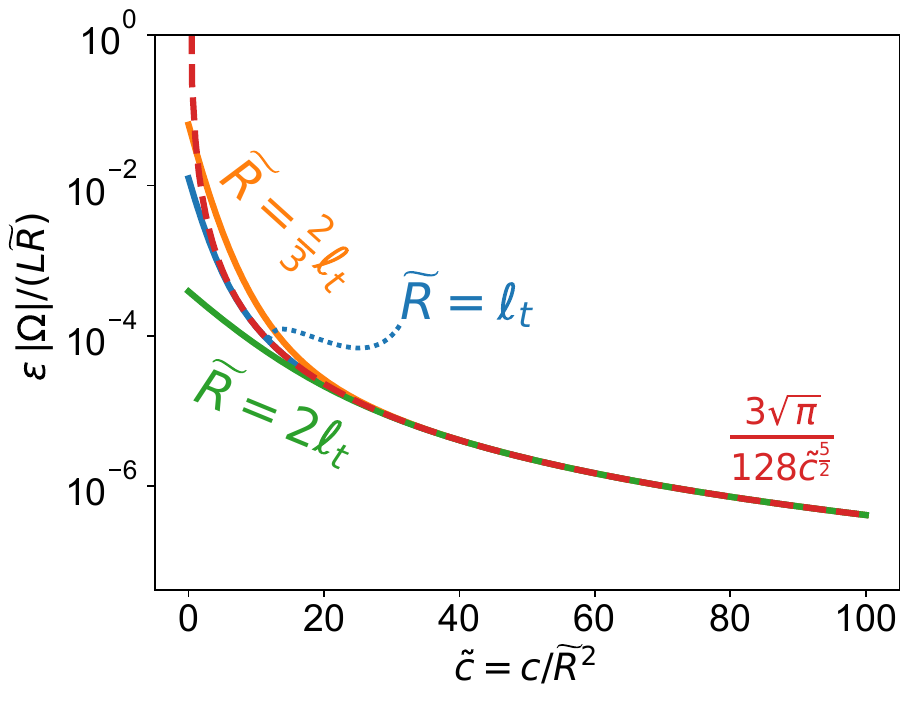} 
\caption{Comparison of the analytical (normalized) threshold $\epsilon\,\tfrac{\abs{\Omega}}{L\Rrad}$ as a function of the (normalized) decay rate $\tilde{c}=c/\Rrad^2$, in 2D, using the conic filter, for the cases $\Rrad=\ell_t$ in blue (\cref{eq:epsilon_conic_exact}), $\Rrad=\tfrac{2}{3}\ell_t$ in orange (\cref{eq:epsilon_alt_1}), and $\Rrad=2\ell_t$ in green (\cref{eq:epsilon_alt_2}), where $\ell_t$ is the target lengthscale. The asymptotic approximation from \cref{eq:epsilon_apprx} is shown as a red dashed line.} 
\label{fig:epsilon_comparison}
\end{figure}

In \cref{sec:param_deriv} we derived an asymptotic approximation of the threshold~$\epsilon$ (\cref{eq:epsilon_apprx}) for the case $\Rrad=\ell_t$, where $\ell_t$ is the target lengthscale. We note, however, that this approximation also holds for cases $\Rrad\approx\ell_t$. We can verify that the exact expression of~$\epsilon$ for the case $\Rrad=\tfrac{2}{3}\ell_t$ is:
\begin{align}
    \epsilon &= \frac{L^{d-1} \Rrad}{\abs{\Omega}}\frac{1}{256 \, \tilde{c}^{5/2}} \Bigg[ 
\sqrt{\pi} (\tilde{c}^2 - 8\tilde{c} + 48) \, \text{erf}(\sqrt{\tilde{c}}) \notag\\
&\quad- \sqrt{\pi} (\tilde{c}^2 - 8\tilde{c} + 42) \, \text{erf}\left(\frac{\sqrt{\tilde{c}}}{2}\right) \notag\\
&\quad- e^{-\tilde{c}/4}(\tilde{c} - 42) \sqrt{\tilde{c}} 
- 48 e^{-\tilde{c}}(\tilde{c} + 2) \sqrt{\tilde{c}} 
\Bigg],
\label{eq:epsilon_alt_1}
\end{align}
whereas for $\Rrad = 2\,\ell_t$ it is:
\begin{align}
    \epsilon &= \frac{L^{d-1} \Rrad}{\abs{\Omega}}\frac{1}{256 \, \tilde{c}^{5/2}} \bigg(
6 \sqrt{\pi} \, \text{erf}\bigg(\frac{\sqrt{\tilde{c}}}{2}\bigg) \notag\\
&\quad- \sqrt{\tilde{c}} \, (\tilde{c} + 6) \, e^{-\tilde{c}/4}
\bigg),
\label{eq:epsilon_alt_2}
\end{align}
where $\tilde{c}=c/\Rrad^2$. All three \cref{eq:epsilon_conic_exact,eq:epsilon_alt_1,eq:epsilon_alt_2} have the same asymptotic approximation for large $\tilde{c}$ given in \cref{eq:epsilon_apprx}, as seen in \cref{fig:epsilon_comparison}. In the discussion at the end of \cref{sec:closed_form_densities_pde} we show that the value of $\epsilon$ is insensitive to the target lengthscale $\ell_t$ using an asymptotic calculation argument. 

\section{Derivations: PDE~filters}\label{sec:closed_form_densities_pde}
In this section we derive the closed-form expressions of the thresholds $\eta_e$ and $\eta_d$, the filtered and projected densities, and the asymptotic value of the geometric constraint threshold~$\epsilon$, when the PDE filtering procedure~\citep{lazarov_filters_2011} is employed. Here, the filtered density $\frho$ is obtained by solving the following modified Helmholtz PDE:
\begin{align}
    \left[-\left(\frac{\Rrad}{2\sqrt{3}}\right)^2\Delta+1\right]\frho = \lrho,
\end{align}
where the parameter $\Rrad$ plays a similar role as the filter radius of the conic filter. The resulting filtered density $\frho$ can be written in terms of the free-space Green's function $\filtk_\Rrad$ as the convolution $\frho = \filtk_\Rrad*\lrho$, where 
\begin{align}\label{eq:green_func_pde}
    \filtk_\Rrad(x) = 
    \begin{cases}
        \frac{\sqrt{3}}{\Rrad}e^{-\frac{2\sqrt{3}}{\Rrad}\abs{x}}, &\text{in 1D,}\\
        \frac{6}{\pi\Rrad^2}K_0\left(\frac{2\sqrt{3}}{\Rrad}\abs{x}\right), &\text{in 2D,}\\
        \frac{3}{\pi\Rrad^2\abs{x}}e^{-\frac{2\sqrt{3}}{\Rrad}\abs{x}}, &\text{in 3D,}
    \end{cases}
\end{align}
where $K_0$ is the modified Bessel function of the second kind of order zero, which satisfies $K_0(r)\sim -\log(r)$ as $r\rightarrow 0^+$ and $K_0(r)\sim \sqrt{\pi/(2r)}e^{-r}$ as $r\rightarrow \infty$, and the normalization ensures that $\int \filtk_\Rrad = 1$. The expression for $\frho$ has an additional contribution coming from the boundary of the design region---due to boundary conditions, usually chosen as homogeneous Neumann boundary conditions---that we can neglect due to the exponential decay of the free-space Green's function.

We now consider the simplified 1D setting presented in \cref{sec:param_deriv} and in \cref{fig:line_density}a, in which the latent density $\lrho$ consists of a 1D binary bump of width $h$: $\lrho(x) = \operatorname{rect}(x/h)$. This latent density is filtered using the 1D Green's function of the PDE filter, $\filtk_\Rrad(x) = \sqrt{3} e^{-\frac{2\sqrt{3}}{\Rrad}\abs{x}}/\Rrad$, resulting in the filtered field:
\begin{align}
\frho(x) &= \left(\filtk_\Rrad*\lrho\right)(x)\\
&= \int_{-\tfrac{\tilde h}{2}}^{\tfrac{\tilde  h}{2}}\sqrt{3}e^{-2\sqrt{3}\abs{\tilde{x}-\tilde{t}}}\de \tilde  t\\
&= 
\begin{cases}
 1 - e^{-\sqrt{3}\tilde{h}} \cosh\left(2\sqrt{3}\abs{\tilde{x}}\right), & \abs{\tilde x}\in\left[0,\frac{\tilde h}{2}\right),\\ \sinh\left(\sqrt{3}\tilde{h}\right)e^{-2\sqrt{3}\abs{\tilde{x}}} , &\abs{\tilde x}\in\left[\frac{\tilde h}{2},\infty\right),
\end{cases}\label{eq:frho_pde}
\end{align}
where we employed normalized units $\tilde x = x/\Rrad$ and $\tilde h = h/\Rrad$. Now, we compute the projected density as a Heaviside projection $(\beta=\infty)$, this is $\prho = H(\frho-1/2)$, which corresponds to a binary bump of width $w$, $\prho(x) = \operatorname{rect}(x/w)$, where the (normalized) width $\tilde w = w/\Rrad$ can be computed in closed-form (by solving $\frho(w/2)=1/2$) as a function of the (normalized) latent width $\tilde h$:
\begin{align}\label{eq:l_to_h_pdefilter}
    \tilde w \left(\tilde h\right) = 
    \begin{cases}
    0, & \tilde{h}\in \left[0,\frac{ \log(2) }{ \sqrt{3}}\right)\\
     \frac{1}{\sqrt{3}} \operatorname{arccosh}\left( \frac{e^{\sqrt{3} \tilde{h}}}{2} \right), & \tilde{h}\in \left[\frac{ \log(2) }{ \sqrt{3}},\infty\right)
    \end{cases}
\end{align}
whereas the the inverse relation is:
\begin{align}\label{eq:h_to_l_pdefilter}
    \tilde h\left(\tilde w\right) = \frac{1}{\sqrt{3}}\log\left(2\cosh\left(\sqrt{3}\tilde{w}\right)\right), \quad \tilde{w}\in (0,\infty).
\end{align}
We now proceed to derive the thresholds $\eta_e$ and $\eta_d$, which differ from the thresholds of the conic filter case given in \cref{eq:eta_e,eq:eta_d}. To compute the threshold $\eta_e$ of the solid constraint~\citep{qian_topological_2013}, we assume that the physical lengthscale $w$ of the feature equals the target lengthscale $\ell_t$ to be imposed---this is $\tilde{w}=w/\Rrad = \ell_t/\Rrad$---and then compute the maximum value of $\frho$, which occurs at $\tilde{x}=0$ and is given by
\begin{align}\label{eq:eta_h_pde}
\tilde{\eta}_e\left(\tilde{h}\right)\coloneq \frho(0) = 1-e^{-\sqrt{3}\tilde{h}},\quad \tilde{h}\in [0,\infty),
\end{align}
where we employed \cref{eq:frho_pde}. The threshold $\eta_e$ is then given by \cref{eq:eta_h_pde} but as a function of the physical width $\tilde{w}$, this is $\eta_e\left(\tilde{w}\right)=\tilde{\eta}_e\left(\tilde{h}\left(\tilde{w}\right)\right)$, where $\tilde{h}\left(\tilde{w}\right)$ is given in \cref{eq:h_to_l_pdefilter}. This yields
\begin{align}\label{eq:eta_e_pdefilter}
    \eta_e\left(\tfrac{\ell_t}{\Rrad}\right) = 1-\frac{1}{2}\operatorname{sech}\left(\sqrt{3}\tfrac{\ell_t}{\Rrad}\right), \quad \tfrac{\ell_t}{\Rrad}\in [0,\infty).
\end{align}
Note that for the typical choice $\Rrad=\ell_t$ we have $\eta_e\approx 0.8284$, which is slightly different from the value $\eta_e= 0.75$ of the conic filter case (\cref{eq:eta_e}). In \cref{fig:eta_comparison} we show a comparison between $\eta_e(\tilde{w})$ for the conic and PDE filter cases. An analogous calculation for the void threshold leads to $\eta_d=1-\eta_e$; explicitly,
\begin{align}\label{eq:eta_d_pdefilter}
    \eta_d\left(\tfrac{\ell_t}{\Rrad}\right) = \frac{1}{2}\operatorname{sech}\left(\sqrt{3}\tfrac{\ell_t}{\Rrad}\right), \quad \tfrac{\ell_t}{\Rrad}\in [0,\infty).
\end{align}
Given the correct thresholds, we proceed to derive~$\epsilon$, where again we assume $\tilde{w}=w/\Rrad = \ell_t/\Rrad$. Unlike the conic filter case, the expression for $\epsilon$ cannot be written in closed-form, such as the one given in \cref{eq:epsilon_conic_exact}; instead, we provide an asymptotic expression, similar to \cref{eq:epsilon_apprx} of the conic filter case, derived as follows. The solid structural function $\bol I^s$ of \cref{eq:is_func} indicates that the largest contribution of the constraint value comes from a neighborhood of $x=0$, the point at which $\frho$ achieves its maximum value. Expanding $\frho$ (\cref{eq:frho_pde}) up to second order around this point yields
\begin{align}
    \frho(x)&= 1 - e^{-\sqrt{3}\tilde{h}} \cosh\left(2\sqrt{3}\abs{\tilde{x}}\right)\\
    &= \underbrace{(1 - e^{-\sqrt{3} \tilde h})}_{\tilde{\eta}_e\left(\tilde{h}\right)} - \underbrace{6 e^{-\sqrt{3} \tilde{h}}}_{\tilde{\gamma}\left(\tilde{h}\right)} \tilde{x}^2 + \mathcal{O}\left(\tilde{x}^4\right) \\
    &= {\eta}_e\left(\tilde{w}\right) - \underbrace{3\operatorname{sech}\left(\sqrt{3}\tilde{w}\right)}_{{\gamma}\left(\tilde{w}\right)} \tilde{x}^2 + \mathcal{O}\left(\tilde{x}^4\right),\label{eq:frho_series_pdefilter}
\end{align}
for $x\approx 0$, where we defined $\tilde{\gamma}\left(\tilde{h}\right)\coloneq -\tfrac{1}{2}\tfrac{\de^2}{\de\tilde{x}^2}\frho(0)=6 e^{-\sqrt{3} \tilde{h}}$ in the second line, and in the last line we replaced $\tilde{h}=\tilde{h}(\tilde{w})$ using \cref{eq:h_to_l_pdefilter} and defined the \emph{correction factor} ${\gamma}\left(\tilde{w}\right)\coloneq\tilde{\gamma}\left(\tilde{h}\left(\tilde{w}\right)\right)$, given by
\begin{align}\label{eq:gamma_pde}
    \gamma\left(\tfrac{\ell_t}{\Rrad}\right)=3\operatorname{sech}\left(\sqrt{3}\tfrac{\ell_t}{\Rrad}\right), \quad \tfrac{\ell_t}{\Rrad}\in [0,\infty),
\end{align}
which will result relevant in the computation of $\epsilon$. 

The value of $\epsilon$ is given by the value of the geometric constraint when $w = \ell_t$ (see \cref{sec:param_deriv}). Starting from \cref{eq:gs_integralform}, we have 
\begin{align}
    \epsilon &= g^s(\lrho,\ell_t)\\
    &=\frac{L^{d-1} \Rrad}{\abs{\Omega}}\int_{-\infty}^{\infty} \bol \prho(\tilde x)e^{-\tilde{c}\abs{\frac{\mathrm{d}}{\mathrm{d} \tilde x}\frho(\tilde x)}^2} \big[\min\{\frho(\tilde x)\notag\\
    &\quad-\eta_e(\tilde{w}), 0\}\big]^2\de \tilde x\\
    &\approx \frac{L^{d-1} \Rrad}{\abs{\Omega}}\int_{-\infty}^{\infty} e^{-4\tilde{c}\left(\gamma(\tilde{w})\right)^2\tilde{x}^2}\left(\gamma(\tilde{w})\right)^2\tilde{x}^4 \de \tilde x\\
    &= \frac{L^{d-1} \Rrad}{\abs{\Omega}} \frac{3\sqrt{\pi}}{128 \tilde c^{\frac{5}{2}}}\frac{1}{\left(\gamma(\tilde{w})\right)^3},
\end{align}
where in the second line we used normalized variables $\tilde x = x/\Rrad$ and $\tilde{c}=c/\Rrad^2$, in the third line we approximated $\frho$ with its second-order expression around $x=0$ (\cref{eq:frho_series_pdefilter}) and $\prho(\tilde{x})=1$, which is valid if $\tilde{c}$ is large since then the most important contribution of the integral comes from a neighborhood around $\tilde x =0$, and finally, we performed the integration in the last line. This yields the approximation
\begin{align}\label{eq:epsilon_apprx_pde}
    \epsilon \approx \frac{L^{d-1} \Rrad}{\abs{\Omega}} \frac{3\sqrt{\pi}}{128 \tilde c^{\frac{5}{2}}}\frac{1}{\left(\gamma\left(\tfrac{\ell_t}{\Rrad}\right)\right)^3},
\end{align}
which is analogous to \cref{eq:epsilon_apprx} for the conic filter case except for a correction factor $\left(\gamma\left(\ell_t/\Rrad\right)\right)^{-3}$, which is close to unity in the typical case $\Rrad=\ell_t$ since $\gamma(1)\approx 1.03$. Note that a similar asymptotic analysis could have been used to derive the expression for $\epsilon$ for the conic filter case in \cref{eq:epsilon_apprx}; it can be shown that this expression does not include a correction factor because $\gamma_\text{conic}\left(\tilde{w}\right) = 1$ for $\tilde{w}<2$, as seen from \cref{eq:rhotilde_1,eq:rhotilde_2}---this explains why, in the conic filter case, the threshold $\epsilon$ is insensitive to the target lengthscale $\ell_t$.

As discussed in \cref{sec:pde_bipde_filters}, we have found that the decay rate $c=10 \Rrad^2$ works well in our numerical examples, which is similar to the value suggested in \citeasnoun{yang_note_2019}. This choice---and the estimates $L/\abs{\Omega}^{\frac{1}{d}}\approx 1$ and $ \Rrad/\abs{\Omega}^{\frac{1}{d}}\approx10^{-2}$ made in \cref{sec:param_deriv}---leads to $\epsilon\approx\left(\gamma\left(\ell_t/\Rrad\right)\right)^{-3}\cdot10^{-6}$ according to \cref{eq:epsilon_apprx_pde}. 

\section{Derivations: bi-PDE~filters}\label{sec:closed_form_densities_bipde}

\begin{figure}[h!]
\centering
\includegraphics[width=0.4\textwidth]{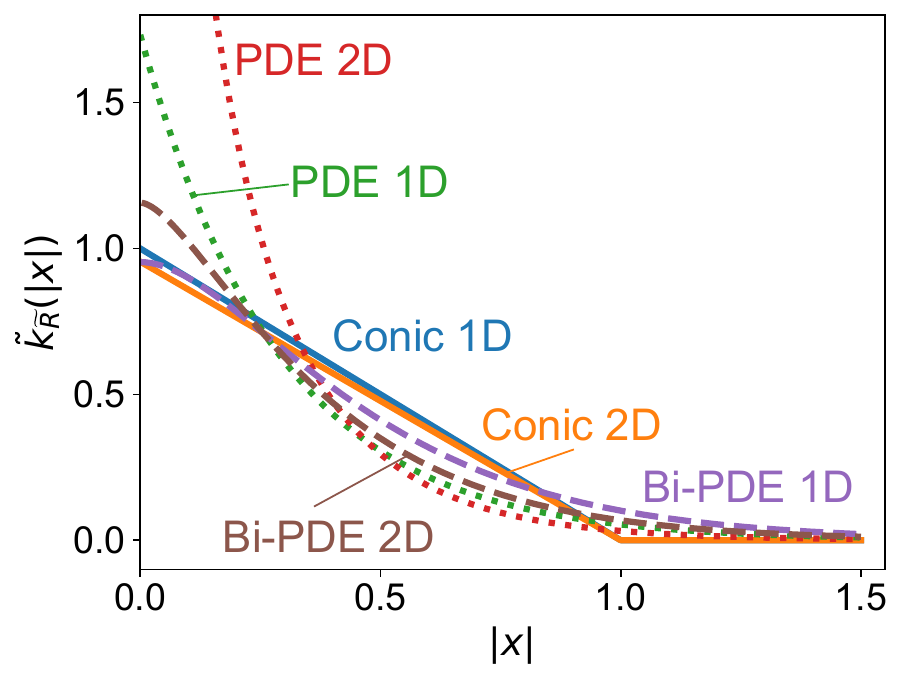} 
\caption{Comparison of the conic filter (\cref{eq:conic_filter}), the PDE filter (\cref{eq:green_func_pde}), and the bi-PDE filter (\cref{eq:green_func_bipde}), as functions of radial distance $\abs{\bol x}$, for $\Rrad = 1$ and dimensions one and two. The 2D PDE filter exhibits a logarithmic singularity at the origin.} 
\label{fig:conic_pde_bipde_filters}
\end{figure}

As discussed in \cref{sec:pde_bipde_filters}, employing the PDE filter results in a filtered density that exhibits large spatial derivatives \correction{$\norm{\nabla_{\nex}\frho(\nex)}$}---caused by the singular PDE Green's function---and this poses challenges to the geometric constraints as well as to robust optimization~\citep{christiansen_creating_2015}. It is desirable to have a PDE filtering scheme that alleviates these issues; this requires a higher-order and rotationally symmetric PDE with a well-behaved Green's function. Analytically, our proposed ``bi-PDE'' filtering scheme consists of obtaining the filtered density $\frho$ by solving the following fourth-order PDE:
\begin{align}
    \left[-\left(r\Rrad\right)^2\Delta+1\right]^2 \frho = \lrho.
\end{align}
This is the composition of two modified Helmholtz PDEs, where $\Rrad$ plays a similar role as the conic filter radius and $r$ is a dimensionless parameter used to adjust the spread of the Green's function. Its Green's function is given by~\citep{lazar2006theory}: 
\begin{align}\label{eq:green_func_bipde}
    \filtk_\Rrad(x) = 
    \begin{cases}
        \frac{  r \Rrad + \lvert x \rvert }{4 r^2 \Rrad^2}e^{- \frac{\lvert x \rvert}{r \Rrad}},
&\text{in 1D,}\\\frac{\lvert x \rvert}{4 \pi r^3 \Rrad^3} K_1\left( \frac{\lvert x \rvert}{r \Rrad} \right)
, &\text{in 2D,}\\
\frac{1}{8\pi r^3 \Rrad^3}e^{- \frac{\lvert x \rvert}{r \Rrad}},
&\text{in 3D,}
    \end{cases}
\end{align}
where $K_1$ is the modified Bessel function of the second kind of order one. We note that it is more regular than the Green's function of the PDE filter (\cref{eq:green_func_pde}): it is finite at the origin---indeed, $\filtk_\Rrad(\bol 0)=\left(4\pi r^2 \Rrad^2\right)^{-1}$ in the 2D case---and all its derivatives are bounded; thus, the spatial derivatives of $\frho$ are bounded as well. \correction{(As discussed in \cref{sec:pde_bipde_filters}, the convolution of two singular PDE Green's functions leads to the smoother bi-PDE Green's function.)} We show a comparison between the conic, PDE and bi-PDE filters in \cref{fig:conic_pde_bipde_filters}. The bi-PDE filter is the double application of the PDE filter of \cref{sec:closed_form_densities_pde} with a different scaling. As such, in practice, the user may reuse existing PDE solver code by first solving $\left[-\left(r\Rrad\right)^2\Delta+1\right] \frho_\text{int} = \lrho$ and then $\left[-\left(r\Rrad\right)^2\Delta+1\right] \frho = \frho_\text{int}$, where $\frho_\text{int}$ is an intermediate variable.  In all of our examples we employed the value $r=r_0$ where
\begin{equation}
r_0\coloneq0.262266719739401 \, ,
\label{eq:r_0}
\end{equation}
which we obtained numerically by minimizing the $L^2$ error over $\R^2$ between the 2D Green's function of \cref{eq:green_func_bipde} and the 2D conic filter of \cref{eq:conic_filter}.  

We now proceed to derive the thresholds $\eta_e$ and $\eta_d$, and the geometric constraint threshold~$\epsilon$, as done in \cref{sec:closed_form_densities_pde}. We consider the simplified 1D setting presented in \cref{sec:param_deriv} and \cref{fig:line_density}a, in which the latent density $\lrho$ is a 1D binary bump of width $h$, $\lrho(x) = \operatorname{rect}(x/h)$, which is filtered using the 1D Green's function of \cref{eq:green_func_bipde}, resulting in the filtered field:
\begin{align}
\frho(x) &= \left(\filtk_\Rrad*\lrho\right)(x) = \\
&
\begin{cases}
\frac{1}{8r} e^{- \frac{\tilde{h} + 2\abs{\tilde{x}}}{2r}} \bigg( -\tilde{h} - 4r +  8 e^{\frac{\tilde{h} + 2\abs{\tilde{x}}}{2r}} r \\
\quad- e^{\frac{2\abs{\tilde{x}}}{r}} (\tilde{h} + 4r - 2\abs{\tilde{x}}) - 2\abs{\tilde{x}} \bigg), \\\quad \text{for } \abs{\tilde x}\in \left[0,\frac{\tilde h}{2}\right),\\
\frac{1}{8r} e^{- \frac{\tilde{h} + 2\abs{\tilde{x}}}{2r}} \bigg( - (1 + e^{\tilde{h}/r}) \tilde{h} \\
\quad+ 2 (-1 + e^{\tilde{h}/r}) (2r + \abs{\tilde{x}}) \bigg), \\\quad \text{for } \abs{\tilde x}\in\left[\frac{\tilde h}{2},\infty\right),
\end{cases}
\end{align}
where we are using normalized units $\tilde x = x/\Rrad$ and $\tilde h = h/\Rrad$. The projected density is computed as a Heaviside projection $(\beta=\infty)$, this is $\prho = H(\frho-1/2)$, resulting in a binary bump of width $w$, $\prho(x) = \operatorname{rect}(x/w)$. There is no closed-form expression for the (normalized) latent width $\tilde{h}$ as a function of the physical width $\tilde{w}$---it can only be obtained numerically. We have found, however, that the expression for $\tilde{h}(\tilde{w})$ of the PDE filter (\cref{eq:h_to_l_pdefilter}) plus a Gaussian correction term is a good approximation when $r=r_0$:
\begin{align}\label{eq:h_to_l_bipdefilter}
    \tilde h\left(\tilde w\right) &\approx \frac{1}{\sqrt{3}}\log\left(2\cosh\left(\sqrt{3}\tilde{w}\right)\right) + Ae^{-B\tilde{w}^2},
\end{align}
for $\tilde{w}\in (0,\infty)$ and $r=r_0$, where $A = 0.197548650630786$ and $B = 1.538127216560406$, which we fitted numerically. This approximation has a maximum absolute error of $10^{-3}$ over the range $\tilde{w}\in (0,2)$, compared to an exact numerical computation.

\begin{figure}[h!]
\centering
\includegraphics[width=0.4\textwidth]{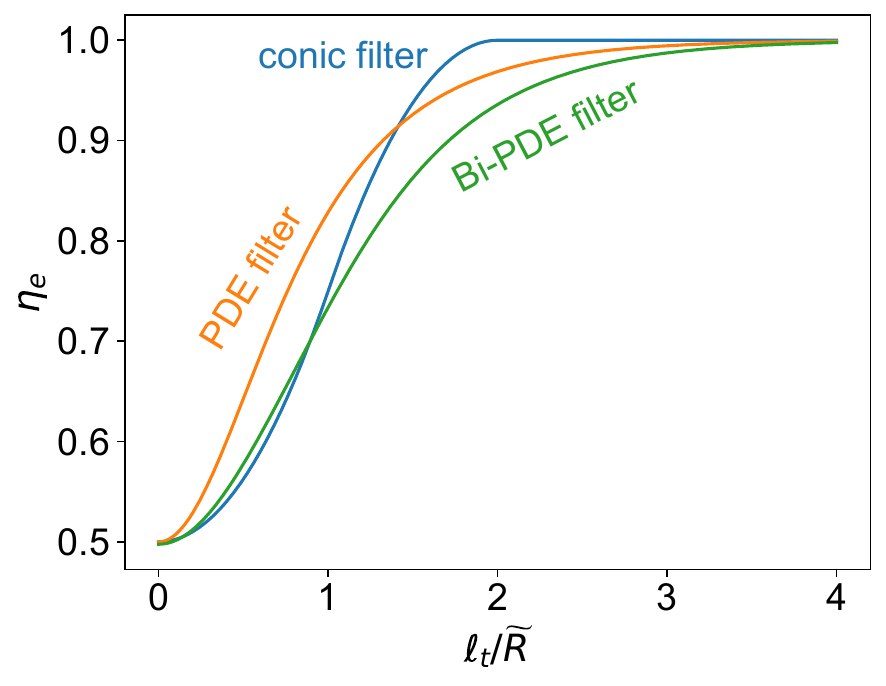} 
\caption{Comparison of the threshold $\eta_e$ for the conic (\cref{eq:eta_e}), PDE (\cref{eq:eta_e_pdefilter}), and bi-PDE (\cref{eq:eta_e_bipdefilter}) filter cases, as a function of (normalized) target lengthscale $\ell_t/\Rrad$.} 
\label{fig:eta_comparison}
\end{figure}

To compute the thresholds $\eta_e$ and $\eta_d$ and the correction factor $\gamma$, we expand $\frho$ to second order around $x=0$:
\begin{align}
    \frho(x)&= \underbrace{-\frac{1}{8r} e^{- \frac{\tilde{h}}{2r}} \left( 2\tilde{h} + 8r - 8 e^{\frac{\tilde{h}}{2r}} r \right)}_{\tilde{\eta}_e\left(\tilde{h}\right)}\notag\\
&\quad-  \underbrace{\frac{1}{8r^3} e^{- \frac{\tilde{h}}{2r}} \tilde{h}}_{\tilde{\gamma}\left(\tilde{h}\right)} \,\tilde{x}^2+ \mathcal{O}\left(\tilde{x}^4\right), \quad x\approx 0,
\label{eq:frho_series_bipdefilter}
\end{align}
hence,
\begin{align}
\tilde{\eta}_e\left(\tilde{h}\right) &= \frho(0) = -\frac{1}{8r} e^{- \frac{\tilde{h}}{2r}} \left( 2\tilde{h} + 8r - 8 e^{\frac{\tilde{h}}{2r}} r \right),\\
\tilde{\gamma}\left(\tilde{h}\right) &=  -\frac{1}{2}\frac{\de^2}{\de\tilde{x}^2}\frho(0)=\frac{1}{8r^3} e^{- \frac{\tilde{h}}{2r}} \tilde{h}.
\end{align}
Then,
\begin{align}
{\eta}_e\left(\frac{\ell_t}{\Rrad}\right) &= \tilde{\eta}_e\left(\tilde{h}\left(\frac{\ell_t}{\Rrad}\right)\right),\label{eq:eta_e_bipdefilter}\\
{\gamma}\left(\frac{\ell_t}{\Rrad}\right) &=\tilde{\gamma}\left(\tilde{h}\left(\frac{\ell_t}{\Rrad}\right)\right), \label{eq:gamma_bipde}
\end{align}
and $\eta_d = 1 - \eta_e$, where $\ell_t$ is the target lengthscale and $\tilde{h}$ is approximated with \cref{eq:h_to_l_bipdefilter} for the case $r=r_0$ from \cref{eq:r_0}. For the typical case $\Rrad=\ell_t$ we have $\eta_e(1) \approx 0.733$ and $\gamma(1) \approx 0.973$, which are similar to the conic filter case values. In \cref{fig:eta_comparison} we show a comparison between $\eta_e(\tilde{w})$ of the conic, PDE, and bi-PDE filters.

Following the same derivation of \cref{sec:closed_form_densities_pde}, the asymptotic expression for the geometric constraint threshold~$\epsilon$ is given in \cref{eq:epsilon_apprx_pde}, with the correction factor $\gamma$ given in \cref{eq:gamma_bipde}. In all our numerical examples we employ the decay rate $c=64\Rrad^2$, the same as in the conic filter case, and the approximation $\epsilon\approx\left(\gamma\left(\ell_t/\Rrad\right)\right)^{-3}\cdot10^{-8}$, obtained with the asymptotic approximation of~$\epsilon$ and the estimates $L/\abs{\Omega}^{\frac{1}{d}}\approx 1$ and $ \Rrad/\abs{\Omega}^{\frac{1}{d}}\approx10^{-2}$ made in \cref{sec:param_deriv}.

\section{Optimization details}\label{sec:unconstr_opt}
In our examples of \cref{sec:results}, in the unconstrained optimization stage, we employed the CCSAQ algorithm to optimize the unconstrained designs. All examples in the same TO challenge (e.g., mode converter) share the same randomly initialized starting design. The filter radius was set equal to the target lengthscale of each example ($\Rrad=\ell_t$). In \cref{tab:unconstr,tab:unconstr_beam_splitter}, we show the $\beta$-scheduling that we used during the unconstrained optimization for each TO challenge---which includes the number of epochs, number of iterations per epoch, and projection steepness parameter $\beta$ used in each epoch. For the cavity example of \cref{sec:cavity}, we employed 25 epochs---each with 10 iterations, with a starting $\beta=24$ that is multiplied by $1.2$ in each epoch---and a final 26th epoch of 20 iterations with $\beta=\infty$. Importantly, in the optimizer settings we set a relative tolerance of $10^{-6}$, which causes some examples to terminate their unconstrained optimization early, before reaching the prescribed number of iterations indicated in this section.
\begin{tablefull}[h!]
            \centering
		\caption{Continuation scheme of $\beta$ used in the unconstrained optimization stage, for the mode converter (\cref{sec:mode_converter}), wavelength demultiplexer  (\cref{sec:wdm}), and heat-transfer problem (\cref{sec:heat_transfer}) examples.} 
            %\scalebox{0.8}{
		\begin{tabular}{llllllllll}
			\toprule
			&                                   & \multicolumn{8}{l}{Unconstrained optimization epoch} \\
			\multicolumn{1}{l|}{Example}                  & \multicolumn{1}{l|}{Parameter}    & 0   & 1   & 2   & 3         & 4  & 5  & 6        & 7 \\ \midrule
			\multicolumn{1}{l|}{Mode converter}           & \multicolumn{1}{l|}{$\beta$}      & 8   & 16  & 30  & $\infty$  &    &    &          &   \\
			\multicolumn{1}{l|}{}                         & \multicolumn{1}{l|}{No. of iter.} & 20  & 20  & 20  & 100       &    &    &          &   \\ \hline
			\multicolumn{1}{l|}{Wavelength demultiplexer} & \multicolumn{1}{l|}{$\beta$}      & 4   & 8   & 16  & 30        & 60 & 90 & $\infty$ &   \\
			\multicolumn{1}{l|}{}                         & \multicolumn{1}{l|}{No. of iter.} & 25  & 25  & 25  & 25        & 50 & 50 & 100      &   \\ \hline
			\multicolumn{1}{l|}{Heat-transfer problem}    & \multicolumn{1}{l|}{$\beta$}      & 8    &  16   &  32   & 64          &    &    &          &   \\
			\multicolumn{1}{l|}{}                         & \multicolumn{1}{l|}{No. of iter.} & 30     &  30   &   30  &  30         &    &    &          &  \\
			\bottomrule
		\end{tabular}
            %}
		\label{tab:unconstr}
	\end{tablefull}
        \begin{tablefull}[h!]
            \centering
		\caption{Continuation scheme of $\beta$ used in the unconstrained optimization stage, for the beam splitter example (\cref{sec:beam_splitter}).}
            %\scalebox{0.8}{
		\begin{tabular}{llllllllll}
			\toprule
			&                                   & \multicolumn{8}{l}{Unconstrained optimization epoch} \\
			\multicolumn{1}{l|}{Beam splitter example}                              & \multicolumn{1}{l|}{Parameter}    & 0   & 1   & 2   & 3  & 4        & 5   & 6        & 7 \\ \midrule
			\multicolumn{1}{l|}{Default}                                            & \multicolumn{1}{l|}{$\beta$}      & 8   & 16  & 30  & 60 & $\infty$ &     &          &   \\
			\multicolumn{1}{l|}{}                                                   & \multicolumn{1}{l|}{No. of iter.} & 20  & 20  & 20  & 20 & 100      &     &          &   \\ \hline
			\multicolumn{1}{l|}{Conic and PDE filters, large lengthscale (16 pix.)} & \multicolumn{1}{l|}{$\beta$}      & 8   & 16  & 30  & 60 & 120      & 240 & $\infty$ &   \\
			\multicolumn{1}{l|}{}                                                   & \multicolumn{1}{l|}{No. of iter.} & 20  & 20  & 20  & 30 & 30       & 30  & 50       &   \\ \hline
			\multicolumn{1}{l|}{PDE filter, medium lengthscale (8 pix.)}            & \multicolumn{1}{l|}{$\beta$}      & 8   & 16  & 30  & 60 & $\infty$ &     &          &   \\
			\multicolumn{1}{l|}{}                                                   & \multicolumn{1}{l|}{No. of iter.} & 20  & 20  & 20  & 20 & 200      &     &          &   \\ \hline
			\multicolumn{1}{l|}{Bi-PDE filter, large lengthscale (16 pix.)}         & \multicolumn{1}{l|}{$\beta$}      & 8   & 16  & 30  & 60 & 120      & 240 & $\infty$ &   \\
			\multicolumn{1}{l|}{}                                                   & \multicolumn{1}{l|}{No. of iter.} & 20  & 20  & 30  & 50 & 50       & 50  & 200      &  \\
			\bottomrule
		\end{tabular}
        %}
		\label{tab:unconstr_beam_splitter}
	\end{tablefull}

\end{appendices}

\bmhead{Supplementary Information}
This work includes supplementary material.

\subsection*{Declarations}
\bmhead{Conflict of interest} The authors declare that they have no conflict of interest.

\bmhead{Funding}
This work was supported in part by the U.S. Army Research Office through the Institute for Soldier Nanotechnologies (Award No. W911NF-23-2-0121) and by the Simons Foundation through the Simons Collaboration on Extreme Wave Phenomena Based on Symmetries.

\bmhead{Ethics approval}
The authors declare that there are no ethical issues involved in this research.

\bmhead{Consent to participate}
The authors declare that they all consented to participate in this research.

\bmhead{Data availability}
All data are included in the article and its supplementary materials.

\bmhead{Replication of results}
Detailed descriptions of the simulations and computational methods are included to ensure reproducibility. The differentiable thermal solver employed in the heat-transfer example is planned for public release in the near future. 

\bmhead{Author Contributions}
Computational simulations and derivations were carried out by Rodrigo Arrieta. Giuseppe Romano was responsible for the numerical simulation of the heat-transfer example. Steven G. Johnson supervised the research. Rodrigo Arrieta prepared the initial draft of the manuscript. All authors contributed to manuscript revision and approved the final version.

\bibliography{biblio}
\end{document}